\documentclass[11pt,a4paper]{article}
\usepackage{jheppub}
\usepackage{amsmath}
\usepackage{amsfonts}
\usepackage{amssymb}
\usepackage[active]{srcltx} 
\usepackage{xspace}

\usepackage{dsfont}

\def\slash#1{\setbox0=\hbox{$#1$}  % set a box for #1
   \dimen0=\wd0     % and get its size
   \setbox1=\hbox{/} \dimen1=\wd1  % get size of /
   \ifdim\dimen0>\dimen1   % #1 is bigger
      \rlap{\hbox to \dimen0{\hfil/\hfil}} % so center / in box
      #1     % and print #1
   \else     % / is bigger
      \rlap{\hbox to \dimen1{\hfil$#1$\hfil}} % so center #1
      /      % and print /
   \fi}      %   
\newcommand{\U}{\mathcal{U}{[\mathcal{C}_{b}]}}
\newcommand{\W}{\mathcal{W}}
 
\newcommand{\Eins}{\mathds{1}}
\newcommand{\strudress}[1]{#1_{XY,Z}^{{\sin}/{\cos}(N\phi_h+\ldots)}}

\newcommand{\nn}{\nonumber \\}
\newcommand{\be}{\begin{equation}}
 \newcommand{\ee}{\end{equation}}
\newcommand{\bea}{\begin{eqnarray}}
\newcommand{\eea}{\end{eqnarray}}
\newcommand{\TMDP}{TMD PDF\xspace}
\newcommand{\TMDF}{TMD FF\xspace}
\newcommand{\TMDPs}{TMD PDFs\xspace}
\newcommand{\TMDFs}{TMD FFs\xspace}
\newcommand{\FFs}{FFs\xspace}
\def\Phperp{\vect{P}_{h\perp}}
\def\pt{\vect{p}_T}
\def\Kt{\vect{K}_T}
\def\bt{\vect{b}_T}
\def\xb{x_{_{\!B}}}
\newcommand{\xbj}{\xb}                   
\newcommand{\zh}{z}
\newcommand{\slim}{\mskip 1.5mu}              
\newcommand{\bm}{\boldsymbol}
\newcommand{\cdott}{{\mskip -1.5mu} \cdot {\mskip -1.5mu}}
\newcommand{\lf}{\left}
\newcommand{\rg}{\right}
\newcommand{\eps}{\epsilon}

\def\ftperp{f_{1T}^{\perp (1) a}}
\def\tildeftperp{\tilde f_{1T}^{\perp (1) a}}

\newcommand{\GammaOp}{\Gamma}
\newcommand{\quark}{\psi}
\newcommand{\nucl}[1]{{#1}}

\newcommand{\elll}{b}
\newcommand{\bra}[1]{\left\langle #1 \right|}
\newcommand{\ket}[1]{\left| #1 \right\rangle}

\newcommand{\vect}[1]{\ensuremath{{\bm{#1}}}}
\newcommand{\vprp}[1]{\vect{#1}_T}
\newcommand{\kei}{{\ensuremath{p}}}
\newcommand{\tcdot}{{\cdot}}
\newcommand{\tAmp}{\ensuremath{\widetilde{A}^{(+)}}}
\newcommand{\tBmp}{\ensuremath{\widetilde{B}^{(+)}}}
\newcommand{\Amp}{\ensuremath{A^{(+)}}}
\newcommand{\Bmp}{\ensuremath{B^{(+)}}}
\newcommand{\mN}{M}
\newcommand{\myeps}{\ensuremath{\epsilon}}
\newcommand{\slfrac}[2]{\left.#1\middle/#2\right.}
\newcommand{\fourint}{  \int \frac{d(\elll \tcdot P)}{(2\pi)} \ e^{ix(\elll \tcdot P)}  \int_0^\infty \frac{d (-\elll^2)}{4\pi} J_0(\sqrt{-\elll^2 \vprp{\kei}^2})\ }
\newcommand{\bpar}{\mathcal{B}_T}
\newcommand{\bparx}{\mathcal{B}_L}
\newcommand{\FTStrufu}{\mathcal{F}}

\newcommand{\hangalign}{\hspace{2cm}&\hspace{-2cm}}
\newcommand{\nminus}{n_-}
\newcommand{\nplus}{n_+}

\title{Bessel-weighted Asymmetries in Semi-Inclusive Deep Inelastic Scattering}

\author[a]{D.~Boer,}
\author[b]{L. Gamberg,}
\author[c]{B.U.~Musch and }
\author[c]{A. Prokudin}

  \affiliation[a]{Theory group, KVI, University of Groningen,\\
  Zernikelaan 25, NL-9747 AA Groningen, The Netherlands}

\affiliation[b]{Division of Science, Penn State University-Berks,\\
  Reading, Pennsylvania 19083, USA}

  \affiliation[c]{Jefferson Lab, \\
12000 Jefferson Avenue, Newport News, VA 23606, USA}

  \emailAdd{d.boer@rug.nl}
  \emailAdd{lpg10@psu.edu}
  \emailAdd{bmusch@jlab.org}
  \emailAdd{prokudin@jlab.org}  

\abstract{The concept of weighted asymmetries is revisited for semi-inclusive deep inelastic scattering. 
We consider the cross section in Fourier space, 
conjugate to the outgoing hadron's transverse momentum, 
where convolutions of transverse momentum dependent parton distribution 
functions  and fragmentation functions become simple products.  Individual asymmetric terms in the cross section can be projected out by means of a
generalized set of weights involving Bessel functions. Advantages of employing these 
Bessel weights are that they suppress (divergent) contributions from high transverse 
momentum and that soft factors cancel in (Bessel-) weighted asymmetries. Also, the resulting 
compact expressions immediately connect to previous work on evolution equations for transverse 
momentum dependent parton distribution and fragmentation functions and to quantities accessible 
in lattice QCD. Bessel-weighted asymmetries are thus model independent observables that augment 
the description and our understanding of correlations of spin and momentum in nucleon structure.}


\begin{document}  

\maketitle

%\pacs{13.88.+e, 13.60.-r, 13.85.Ni}

\section{Introduction}
\label{sec-intro}
In the factorized picture of semi-inclusive processes,  
where the transverse momentum of the detected hadron  $\Phperp$ is small compared to the photon virtuality $Q^2$,  
transverse momentum dependent (TMD) parton distribution functions (PDFs) characterize 
the spin and momentum structure of the proton 
\cite{Ralston:1979ys,Sivers:1989cc,Collins:1992kk,
Kotzinian:1994dv,Mulders:1995dh,Kotzinian:1997wt,Boer:1997nt}.
At leading twist there are 8 \TMDPs. 
They can be studied experimentally by analyzing angular modulations 
in the differential cross section, so called 
spin and azimuthal asymmetries. These modulations are a function of the azimuthal angles of 
the final state hadron momentum about
the virtual photon direction,  as well as that of the target polarization (see e.g., Ref.~\cite{Bacchetta:2006tn} for a review). 
 \TMDPs enter the SIDIS cross section in momentum space 
convoluted with transverse momentum dependent fragmentation functions (\TMDFs).  However, after a two-dimensional Fourier transform of the cross section with respect to the transverse hadron momentum $\Phperp$,
these convolutions become simple products of functions in Fourier $\bt$-space. 
The usefulness of Fourier-Bessel transforms in 
studying the factorization as well as the scale dependence of 
transverse momentum dependent cross section
has been known for some time~\cite{Parisi:1979se,Jones:1980my,Collins:1981uw,Ellis:1981sj,
Collins:1984kg,Ji:2004xq,CollinsBook}.   
In this paper we exhibit the  structure of the cross section in  $\bt$-space
and demonstrate how this representation results in 
model independent observables which are generalizations of 
the conventional weighted asymmetries~\cite{Kotzinian:1997wt,Boer:1997nt}.
Further we explore the impact that these observables have in studying
the scale dependence of the SIDIS cross section at small to  moderate 
transverse momentum where the TMD framework is designed to give a good 
description of the cross section.  In particular we study how the so called
soft factor cancels from these observables. 
The soft factor~\cite{Collins:1999dz,CollinsBook,Collins:2004nx,Ji:2004xq,Ji:2004wu,Aybat:2011zv} is an essential  element of the cross section  that emerges in the proofs of  TMD factorization~\cite{Collins:1981uw,Collins:1984kg,Ji:2004xq,CollinsBook}. It 
accounts for  the collective effect of soft momentum gluons not 
associated with either the distribution or fragmentation part of the  process 
and it is shown to be universal in hard processes \cite{Collins:2004nx}.  Depending on the factorization framework, it appears explicitly in 
the structure functions and thus in the factorized cross section (see Refs. \cite{Ji:2004xq,Ji:2004wu}), or 
it is completely absorbed in the definition of \TMDPs and \TMDFs (see Refs. \cite{CollinsBook,Aybat:2011zv}).  At tree level (zeroth order in $\alpha_S$) the soft factor is unity, which explains its absence in the factorization formalism considered for example in Ref.~\cite{Bacchetta:2006tn}.
However, for a correct 
description of the energy scale dependence of the cross sections
and asymmetries involving \TMDPs, it is essential to include 
the soft factor.  Yet, it is possible to consider observables 
where the soft factor is indeed  absent or cancels out, 
these are precisely the weighted asymmetries.  

\subsection{Overview on weighted asymmetries}
\label{weights}
The concept of transverse momentum weighted single spin asymmetries (SSA) was proposed 
some time ago in  Refs.~\cite{Kotzinian:1997wt,Boer:1997nt}.  
Using the technique of weighting enables one 
to disentangle in a model independent way the cross sections 
and asymmetries in terms of the transverse (momentum) moments of \TMDPs. A comprehensive  
list of such weights was derived in Ref.~\cite{Boer:1997nt} for 
semi-inclusive deep inelastic scattering (SIDIS).
A prominent example is the weighted Sivers asymmetry, 
obtained from the differential cross section $d\sigma$ according to
%\begin{widetext}
\bea
	A^{w_1 \sin(\phi_h-\phi_S)}_{UT,T} & =& \nn  
&&	\hspace{-2cm}2\;\frac{\int d |\Phperp|\, |\Phperp|d \phi_h\, d \phi_S\, w_1(|\Phperp|)\ \sin(\phi_h - \phi_S)\ \big\{ d\sigma(\phi_h,\phi_S) - d\sigma(\phi_h,\phi_S+\pi )\big\} }
	            {\int d |\Phperp|\, d \phi_h\, |\Phperp| d \phi_S\,  w_0(|\Phperp|)\  \big\{ d\sigma(\phi_h,\phi_S) + d\sigma(\phi_h,\phi_S+\pi ) \big\}},
	\label{eq:BMwtasym}
\nn
\eea
%\end{align}
%\end{widetext}
where the integrations are performed over the observed transverse hadron momentum $|\Phperp|$, the hadron azimuthal angle $\phi_h$ and the spin direction $\phi_S$ of the transversely polarized target, and  the  weights are $w_1 =  |\Phperp|/\zh M$, $w_0=1$. At tree level  and leading twist the weighted 
Sivers asymmetry~\cite{Boer:1997nt} then becomes,  
\begin{align}
A_{UT}^{\frac{|\Phperp|}{\zh M}\sin(\phi_h - \phi_s)} = &  -2  \frac{\sum_a e_a^2\  \ftperp(x) \ D_1^{(0)a}(z) }{ \sum_a e_a^2\ f_{1}^{(0)a}(x)\ D_{1}^{(0)a}(z) }\; ,
\label{eq:ssa_sivers_finalbpar0}
\end{align}
where $f_{1T}^{\perp(1)a}$, $f_{1}^{(0)a}$ and $D_{1}^{(0)a}$ are transverse momentum moments  of \TMDPs  and \TMDFs , and $e_a$ is the electric charge for a quark of flavor $a$.  As explained in greater detail in Section \ref{sec-weightedasym},  the moments in Eq.\ \eqref{eq:ssa_sivers_finalbpar0} are undefined without a subtraction prescription for the infinite contributions at large transverse momentum.
Here, we propose generalized weights, $w_n\propto J_n( |\Phperp| \mathcal{B}_T )$
with $J_n$ denoting Bessel functions of the first kind, and where 
 $\mathcal{B}_T$ (in units $(\mathrm{GeV}/c)^{-1}$) is a free parameter that represents the Fourier conjugate to $|\Phperp|$. For the Sivers asymmetry, 
 $w_1= 2 J_1( |\Phperp| \mathcal{B}_T ) / z M \mathcal{B}_T $ and 
$w_0 = J_0( |\Phperp| \mathcal{B}_T )$. This gives rise to the Bessel-weighted Sivers asymmetry, which reads
\begin{align}
  A_{UT}^{\frac{{2 J}_1(|\Phperp|\bpar )}{\zh M \bpar} \sin(\phi_h - \phi_s)}
 & =\  -2 \frac{\sum_a e_a^2\,  \tildeftperp (x,z^2 \bpar^2)\,  \tilde D_1^{(0)a}(z,\bpar^2) }{ \sum_a e_a^2\, \tilde f_{1}^{(0)a}(x,z^2 \bpar^2)\, \tilde D_{1}^{(0)a}(z,\bpar^2) },
\label{eq:ssa_sivers_final_intro}
\end{align}
where $\tildeftperp$, $\tilde f_{1}^{(0)a}$ and $\tilde D_1^{(0)a}$ are \TMDPs  and \TMDFs  
Fourier transformed with respect to transverse momentum as defined in the next section.
In the asymptotic limit $\mathcal{B}_T \rightarrow 0$, we recover the conventional
 weighted asymmetry 
Eq.\ \eqref{eq:ssa_sivers_finalbpar0}, and the Fourier transformed \TMDPs and FFs can be identified with the moments in that equation. 
An advantage of the generalized weight relates to the asymptotic behavior of \TMDPs (and \TMDFs).
We will see that this provides a regularization  of the infinite contributions at large transverse 
momentum as long as we keep $\bpar^2$ non-zero. 
Moreover, our analysis will show that soft factors appearing beyond tree level 
  cancel out of  the weighted asymmetry.

The rest of the manuscript is organized as follows: In Section \ref{sec-crosssec} we write down the general form of the SIDIS cross section 
in the TMD factorization framework and show that the convolutions in momentum space appear as 
products in Fourier space. For simplicity,
this discussion is presented  at tree level. Modifications needed to go beyond tree level are discussed in Section \ref{sec-beyondtree}.
Even though our arguments are quite general, for definiteness 
we use the framework of Ji, Ma, Yuan~\cite{Ji:2004xq,Ji:2004wu}, 
here referred  to as the ``JMY'' framework. 
\TMDPs at the level of matrix elements will be considered in Section~\ref{sec-matrixel}.
In Section \ref{sec-weightedasym} we will consider Bessel-weighted asymmetries, focusing on the Sivers asymmetry as an explicit example.
Further asymmetries at leading twist are listed in Appendix \ref{sec-bwa}. 
We will also consider $x$ moments of \TMDPs and introduce
a method to study Fourier transformed moments in lattice QCD and 
compare with experiment. 
Our conclusions are presented in  Section \ref{sec-conl}.

\section{The SIDIS cross section in Fourier space at tree level}
\label{sec-crosssec}

\subsection{Elements of the SIDIS cross section}
 
The lepton-hadron cross section of SIDIS 
$\ell(l) + N(P,S) \rightarrow \ell(l) + h(P_h) + X$ can
be expressed \cite{Gourdin:1973qx,Kotzinian:1994dv,Diehl:2005pc,Bacchetta:2006tn} in the 
notation of Ref.~\cite{Bacchetta:2006tn} as
\bea
\frac{d\sigma}{d\xbj \, dy\, d\psi \,dz_h\, d\phi_h\, |\Phperp|\,d |\Phperp|}
&=&
\frac{\alpha^2}{\xbj y\slim Q^2}\,
\frac{y^2}{(1-\varepsilon)}\,  \biggl( 1+\frac{\gamma^2}{2\xbj} \biggr)\,
L_{\mu \nu} W^{\mu \nu},
\label{eq:crossmaster}
\eea
where we assume one photon exchange. 
\begin{figure}
\begin{center}
	\includegraphics[width=0.5\textwidth]{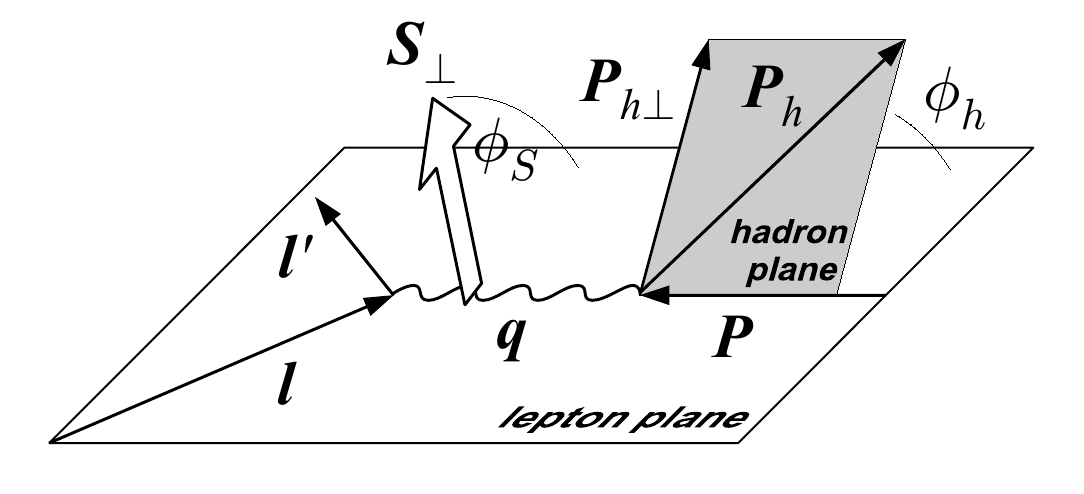}
	\caption{Kinematics of the SIDIS process, compare Refs.\ \cite{Bacchetta:2004jz,Bacchetta:2006tn}.\label{fig-kinem}}
\end{center}
\end{figure}
$L_{\mu\nu}$ and $W^{\mu\nu}$ are the leptonic and hadronic tensors respectively,  and the vector $\Phperp$ is the transverse momentum of the produced hadron 
in a frame where the virtual photon and the target are collinear, e.g.\ in the target rest frame or $\gamma^* P$ center of mass frame. It makes an azimuthal angle $\phi_h$ with the lepton scattering plane defined by the momenta of the incoming and the final leptons $l$ and $l'$ (see Figure~\ref{fig-kinem}). We define $q \equiv l-l'$, and $q^2 = -Q^2$ is the virtuality of the photon. $\psi$ is the azimuthal angle of $l'$ around the lepton beam axis relative to $S_\perp$, in DIS kinematics $d\psi \approx d\phi_S$~\cite{Diehl:2005pc}.  The subscript ``$_{\perp}$'' denotes transverse projection in the target rest frame while the subscript ``$_{ T}$" denotes transverse projection in the light-cone frame. We use definitions for the kinematic variables and the ratio of  longitudinal and transverse photon flux $\epsilon$ as in Ref. \cite{Bacchetta:2006tn},
\be
\xbj = \frac{Q^2}{2\,P\cdott q}, \;\;
y = \frac{P \cdott q}{P \cdott l}, \;\;
z_h = \frac{P \cdott P_h}{P\cdott q}, \;\;
\gamma = \frac{2 M x}{Q}, \;\; \varepsilon = \frac{1-y -\frac{1}{4}\slim \gamma^2 y^2}{1-y
  +\frac{1}{2}\slim y^2 +\frac{1}{4}\slim \gamma^2 y^2} \ ,
  \label{eq:xyz}
\ee
where $M$ is the mass of the target nucleon.  
We employ the standard light-cone decomposition of four-vectors $\omega^\mu = \omega^+ \nplus^\mu + \omega^- \nminus^\mu + \omega_T^\mu$. In the $\gamma^*  P$ center of mass frame with the proton three-momentum pointing in positive $z$-direction, the nucleon carries no transverse momentum, $P_T=0$, and $x \equiv p^+/P^+$ denotes the momentum fraction carried by the quark (parton) of momentum $p$.   Further definitions of kinematic variables  and 
details on the leptonic and hadronic
tensor are  given in Appendix~\ref{sec-convent} and  Ref.~\cite{Bacchetta:2006tn}.

At tree-level of the hard photon-quark scattering process, and to  leading 
order in the $1/Q$ expansion, the hadronic tensor can be written in factorized form as~\cite{Mulders:1995dh,Boer:2003cm,Bacchetta:2006tn}
\be
 2 M W^{\mu\nu}= \sum_a e_a^2 \int d^2\pt\, d^2\Kt\, \delta^{(2)}(z \pt + \Kt-\Phperp) \, {\rm Tr}\Big\{
\Phi(x,\pt )  \gamma^\mu \Delta(z,\Kt ) \gamma^\nu 
\Big\}\, .
\label{eq:hadronic_tensor_momentum}
\ee
The  quark-quark correlator~\cite{Soper:1979fq,Collins:1981uw} 
in the above equation is defined as 
\bea
 \Phi_{ij} 
(p,P,S) & \equiv\ &  \int 
        \frac{d^4 b}{(2\pi)^{4}}\;  e^{i p \cdot b}\,
       \langle P, S|\overline{\psi}_j(0)\,
{\U} 
\psi_i(b)|P ,S\rangle\, .
\eea
In Eq.~\eqref{eq:hadronic_tensor_momentum}  an integration 
has been performed over the 
small component $p^-$ of parton momentum to obtain a correlator that depends
on light-cone fraction $x$ and parton transverse momentum $\pt$, that is,
\bea
\Phi_{ij}(x,\pt) & \equiv\ & \int d p^- {\Phi_{ij}} 
(p,P,S)\nn
 &=& \int 
        \frac{d b^- d^2 \bt}{(2\pi)^{3}}\; 
 e^{i x P^+ b^- - i \pt \cdot \bt}\,
       \langle P, S|\overline{\psi}_j(0)\,
{\U} 
\psi_i(b)|P ,S\rangle \bigg|_{b^+=0}\, .
\label{eq:phi} 
\eea
The state $\ket{\nucl{P,S}}$ represents a nucleon with four-momentum $P$ and spin polarization vector $S$,
and quark fields are located at position ``$0$'' and ``$b$'' in coordinate space. The gauge link 
$\U$ ensures
gauge invariance of the correlator \cite{Boer:2003cm,Belitsky:2002sm}. It corresponds
to a path in $b$ space which  is determined
by the color flow in the hard sub-process~\cite{Bomhof:2004aw,Bomhof:2006dp}. 
We will discuss the details of the definition of the correlator and the 
role of the gauge link
$\U$ in Section ~\ref{sec-matrixel}. Analogous expressions define the fragmentation correlator $\Delta_{ij}(z,\pt)$ (see e.g. \cite{Bacchetta:2006tn}). 

\subsection{Representation in Fourier space}\label{sec:FT}
 
In this section, we rewrite the SIDIS cross section and its transverse momentum dependent components in coordinate $\bt$ space, similar as previously done in Ref.~\cite{Idilbi:2004vb}. Here however, we take advantage of the rotational invariance of \TMDPs and FFs.

First we use the representation of the $\delta$-function
\begin{equation}
	\delta^{(2)}(z \pt + \Kt- \Phperp) = \int \frac{d^2\bf \bt}{(2\pi)^2}\, e^{i \bt (z \pt + \Kt- \vect{P}_{h \perp})}\, ,
\end{equation}
along with the following definitions,
\begin{align}
 W^{\mu \nu}(\Phperp) & \equiv  \int \frac{d^2\bt}{(2\pi)^2}\, e^{-i \bt\cdot \vect{P}_{h \perp}}\; \tilde W^{\mu \nu}(\bt) \, ,\\
\tilde \Phi_{ij}(x, z \bt) & \equiv \int  d^2\pt \, e^{i z \bt \cdot\pt }\; \Phi_{ij}(x, \pt)\nn 
	& = 
	 \int \frac{d b^-}{(2\pi)}\;  e^{i x P^+ b^-}\, 
	 \langle P, S|\overline{\psi}_j(0)\, {\U} \psi_i(b)|P ,S\rangle \bigg|_{b^+=0}\, ,
	\label{eq:phi1}  \\
\tilde \Delta_{ij}(z, \bt) & \equiv \int d^2\Kt \, e^{i   \bt\cdot\Kt }\; \Delta_{ij}(z, \Kt) \, , 
\end{align}
to re-write  the leading term in the 
hadronic tensor, Eq.~\eqref{eq:hadronic_tensor_momentum}, in  
Fourier space
\begin{equation}
 2 M  \tilde W^{\mu\nu} = \sum_a e_a^2 \, {\rm Tr}\left(\tilde \Phi(x, z \bt) \gamma^\mu \tilde \Delta(z, \bt) \gamma^\nu\right) \, .
\label{eq:hadronic_tensor_bspace}
\end{equation}
The advantage of the $\bt$ space representation is clear: the hadronic tensor 
is no longer a convolution of $\pt$ and $\Kt$ dependent functions but a simple product of $\bt$-dependent functions. This  motivates us to re-write the entire cross section in terms of the Fourier transform
\be
\frac{d\sigma}{d\xbj \, dy\, d\psi \,dz_h\, d\phi_h\, |\Phperp|d |\Phperp|} = \int \frac{d^2 \vect{b}_{T}}{(2\pi)^2} e^{-i \bt \cdot \Phperp}  \left\{  \frac{\alpha^2}{\xbj y\slim Q^2}
\frac{y^2}{(1-\varepsilon)}  \biggl( 1+\frac{\gamma^2}{2\xbj} \biggr)  L_{\mu\nu}  \tilde W^{\mu\nu} \right\} .
\end{equation}
Next, we decompose the correlators $\tilde \Phi$ and $\tilde \Delta$ into \TMDPs and \FFs in Fourier space.  Using the trace notation (see also Eqs.\
\eqref{eq-untracephi} and \eqref{eq-untracedelta} in the appendix)
\begin{equation}
 \tilde \Phi^{[\Gamma]} \equiv \frac{1}{2}{\rm Tr} (\tilde \Phi \Gamma) \, ,
\label{eq:bspace_trace}
\end{equation}
and restricting ourselves to leading twist projections, we obtain the following structures for $\tilde \Phi$
 \begin{eqnarray} 
\tilde \Phi^{[\gamma^+]}(x,\bt) & = &
 \tilde f_1
(x,\bt^{2}) 
 - i\, \eps_{T}^{\rho\sigma} b_{T\rho}^{}S_{T \sigma}^{}  \, 
   M \tilde f_{1T}^{\perp(1)}
(x,\bt^{2}) 
 \,,
\nonumber \\
\tilde \Phi^{[\gamma^+ \gamma^5]}(x, \bt) & = &
 S_L \, \tilde g_{1L}
(x,\bt^{2}) 
 + i \, \bt \cdott \vect{S}_T M \, 
   \tilde g_{1T}^{(1)}
(x,\bt^{2}) 
 \,,
\nonumber \\ 
\tilde \Phi^{[i \sigma^{\alpha +}\gamma^5]}(x,\bt) & = &
 S_{T}^{\alpha} \, \tilde h_{1}
(x,\bt^{2}) 
 + i\, S_L\,{b^{\alpha}_T}{M} \, \tilde h_{1L}^{\perp (1)}
(x,\bt^{2}) 
\nonumber \\   & & %\hskip -0.4cm
 + {\frac{1}{2}\left(b_T^{\alpha} b_T^{\rho}
     +\frac{1}{2}\,\bt^{2}\,g_T^{\alpha\rho}\right)}{M^2}\, S_{T \rho} 
   \tilde h_{1T}^{\perp (2)}
(x,\bt^{2}) \nn & & %\hskip -0.4cm
- i\, {\eps_{T}^{\alpha\rho} b_{T \rho}^{}}{M}  
   \tilde h_{1}^{\perp (1)} 
(x,\bt^{2}) \, ,
\label{eq:correlator_bspace}
\end{eqnarray}
 where $\alpha = 1,2$ and  $\rho=1,2 $. Similarly, we obtain the following structures for $\tilde \Delta$
\begin{eqnarray} 
\tilde \Delta^{[\gamma^-]}(z,\bt) & = &
 \tilde D_1
(z,\bt^{2}) 
 -  i\, \eps_{T}^{\rho\sigma} b_{T \rho}^{}S_{hT \sigma}^{}  \, 
   z M_h \tilde D_{1T}^{\perp(1)}
(x,\bt^{2}) 
 \,,
\nonumber \\
\tilde \Delta^{[\gamma^- \gamma^5]}(z, \bt) & = &
 S_{hL} \, \tilde G_{1L}
(z,\bt^{2}) 
 - i \, \bt \cdott \vect{S}_{hT}{z M_h} \, 
   \tilde G_{1T}^{(1)}
(z,\bt^{2}) 
 \,,
\nonumber \\ 
\tilde \Delta^{[i \sigma^{\alpha -}\gamma^5]}(z,\bt) & = &
 S_{hT}^{\alpha} \, \tilde H_{1}
(z,\bt^{2}) 
 - i\, S_{hL}\,{b^{\alpha}}{z M_h} \, \tilde H_{1L}^{\perp (1)}
(z,\bt^{2}) 
\nonumber \\   & & %\hskip -0.4cm
 + {\frac{1}{2} \left(b_T^{\alpha} b_T^{\rho}
     +\frac{1}{2}\,\bt^{2}\,g_T^{\alpha\rho}\right)}{z^2 M_h^2}\, S_{hT \rho} 
   \tilde H_{1T}^{\perp (2)}
(z,\bt^{2}) \\   & & %\hskip -0.4cm
- i\, {\eps_{T}^{\alpha\rho} b_{T \rho}^{}}{z M_h}  
\tilde H_{1}^{\perp (1)} 
(z,\bt^{2}) \, .
\label{eq:correlator_ff_bspace}
\end{eqnarray}
For future applications, we have written down the latter decomposition for the more general case of a spin-$\frac{1}{2}$ hadron;
the expression for a spinless hadron is obtained by setting $S_h=0$. 
The above decompositions can be deduced from the existing expressions for $\Phi$ and $\Delta$ in momentum space \cite{Mulders:1995dh,Goeke:2005hb}, or starting from the symmetry properties of the correlators $\tilde \Phi$ and $\tilde \Delta$ and a parameterization in terms of Lorentz-invariant amplitudes, see also Section \ref{sec-matrixel} and Appendix \ref{sec-amppar}. 
The functions $\tilde f_1(x,\bt^{2})$, $\tilde g_{1L}(x,\bt^{2})$, $\ldots$ are the Fourier transforms of the usual 
\TMDPs $f_1(x,\pt^{2})$, $g_{1L}(x,\pt^{2})$, $\ldots$.  For a  generic \TMDP called $f$ and a generic \TMDF called $D$, this Fourier transform is given by
\begin{align}
\tilde f(x, \bt^2) & \equiv \int  d^2 \pt \, e^{i \bt\cdot \pt }\; f(x, \pt^2)\nn
	&= 2\pi \int      d|\pt| |\pt|\ J_0(|\bt||\pt|)\  f(x, {\pt^2} )\ , \\
\tilde D(z, \bt^2) & \equiv \int d^2 \Kt \,  e^{i  \bt\cdot \Kt }\; D(z, \Kt^2) 
	= 2\pi\int  d|\Kt| |\Kt|\ J_0(|\bt||\Kt|)\  D(z, {\Kt^2} )\; .
	\label{eq-bttransf}
\end{align}
Additionally, in Eqs.~\eqref{eq:correlator_bspace} and \eqref{eq:correlator_ff_bspace} not only Fourier transformed \TMDPs and \TMDFs, but
also their $\bt^2$-derivatives appear, which we denote as
\begin{align}
	\tilde f^{(n)}(x, \bt^2 ) & \equiv n!\left( -\frac{2}{M^2}\partial_{\bt^2} \right)^n \ \tilde f(x, \bt^2 ) \nn
	&= \frac{ 2\pi \ n!}{(M^2)^n} \int  d |\pt| |\pt|\left( \frac{|\pt|}{|\bt|}\right)^n J_n(|\bt||\pt|)\  f(x, {\pt^2} )\ , \\
	\tilde D^{(n)}(z, \bt^2 )  & \equiv n!\left( -\frac{2}{z^2 M_h^2} \partial_{\bt^2}\right)^n \tilde D(z, \bt^2 )\nn
	&= \frac{ 2\pi \ n!}{(z^2 M_h^2)^n} \int  d |\Kt| |\Kt|\left( \frac{|\Kt|}{|\bt|}\right)^n J_n(|\bt||\Kt|)\  D(z, {\Kt^2} )  \ .
 \label{eq-btder}
\end{align}
The functions $\tilde f$, $\tilde D$, $\tilde f^{(n)}$ and $\tilde D^{(n)}$ are real valued and $\tilde f^{(0)} = \tilde f$, $\tilde D^{(0)} = \tilde D$.
Taking the  ``asymptotic limit'' $|\bt|\rightarrow 0$ on the right hand side of Eqs.~\eqref{eq-btder}, we formally obtain the conventional moments of the \TMDPs and \TMDFs, $f^{(n)}(x)$ and $D^{(n)}(z)$ respectively,
\begin{align}
	\tilde f^{(n)}(x, 0 ) & =   \int  d^2 \pt \left( \frac{\pt^2}{2 M^2}\right)^n    f(x, \pt^2 ) \equiv  f^{(n)}(x)\ , \quad \nonumber \\
	\tilde D^{(n)}(z, 0 ) & =   \int  d^2\Kt \left( \frac{\Kt^2}{2 z^2 M_h^2}\right)^n    D(x, \Kt^2 ) \equiv  D^{(n)}(z)  .
 \label{eq-btder2}
\end{align}
Thus  we find that the derivatives in $\bt$-space are directly related to moments of \TMDPs and \FFs.  Finally we re-write the SIDIS cross section of Ref.~\cite{Bacchetta:2006tn} 
in the  $\gamma^* P$ center of mass frame with the proton three-momentum pointing in the negative $z$-direction (so called Trento conventions \cite{Bacchetta:2004jz}),
as

\begin{align}
\lefteqn{\frac{d\sigma}{d\xbj \, dy\, d\phi_S \,dz_h\, d\phi_h\, |\Phperp|d |\Phperp|} = } 
\nonumber \\   & \quad
\frac{\alpha^2}{\xbj y\slim Q^2}\,
\frac{y^2}{(1-\varepsilon)}\,  \biggl( 1+\frac{\gamma^2}{2\xbj} \biggr)\, \int \frac{d |\bt|}{(2\pi)} |\bt|\Biggl\{
J_{0} (|\bt| |\Phperp|)\, \FTStrufu_{UU,T}  
+ 
\varepsilon\slim
J_{0} (|\bt| |\Phperp|)\, \FTStrufu_{UU ,L}
\nonumber \\  & \quad
+ \quad \sqrt{2\,\varepsilon (1+\varepsilon)}\,\cos\phi_h\,
J_{1} (|\bt| |\Phperp|)\,\FTStrufu_{UU}^{\cos\phi_h}
 + 
\varepsilon \cos(2\phi_h)\, 
J_{2} (|\bt| |\Phperp|)\, \FTStrufu_{UU}^{\cos(2\phi_h)} 
\nonumber \\  & \quad
+ \lambda_e\, \sqrt{2\,\varepsilon (1-\varepsilon)}\, 
           \sin\phi_h\, 
J_{1} (|\bt| |\Phperp|)\,\FTStrufu_{LU}^{\sin\phi_h}
\phantom{\Bigg[ \Bigg] }
\nonumber \\  & \quad 
+ \quad S_\parallel\, \Bigg[
\sqrt{2\, \varepsilon (1+\varepsilon)}\,
  \sin\phi_h\, 
J_{1} (|\bt| |\Phperp|)\,\FTStrufu_{UL}^{\sin\phi_h}
+  \varepsilon \sin(2\phi_h)\, 
J_{2} (|\bt| |\Phperp|)\,\FTStrufu_{UL}^{\sin 2\phi_h}
\Bigg]
\nonumber \\  &  \quad
+ S_\parallel \lambda_e\, \Bigg[ \,
  \sqrt{1-\varepsilon^2}\,
J_{0} (|\bt| |\Phperp|)\, \FTStrufu_{LL}
+\sqrt{2\,\varepsilon (1-\varepsilon)}\,
  \cos\phi_h\, 
J_{1} (|\bt| |\Phperp|)\, \FTStrufu_{LL}^{\cos\phi_h} 
\Bigg]
\nonumber \\ &  \quad
 +
|\bm{S}_\perp|\, \Bigg[
  \sin(\phi_h-\phi_S)\, J_{1} (|\bt| |\Phperp|)\, \,
\Bigl(\FTStrufu_{UT,T}^{\sin(\phi_h-\phi_S)}  
+ \varepsilon\, \FTStrufu_{UT ,L}^{\sin\lf(\phi_h -\phi_S\rg)}\Bigr)
\nonumber \\ &  \quad \qquad
+ 
\; \varepsilon\, \sin(\phi_h+\phi_S)\, 
J_{1} (|\bt| |\Phperp|)\, \FTStrufu_{UT}^{\sin(\phi_h+\phi_S)}
\nonumber \\  & \quad \qquad 
+
\; \varepsilon\, \sin(3\phi_h-\phi_S)\,
J_{3} (|\bt| |\Phperp|)\, \FTStrufu_{UT}^{\sin\lf(3\phi_h -\phi_S\rg)}
\phantom{\Bigg[ \Bigg] }
\nonumber \\  & \quad \qquad 
+
\; \sqrt{2\,\varepsilon (1+\varepsilon)}\, 
  \sin\phi_S\, 
J_{0} (|\bt| |\Phperp|)\, \FTStrufu_{UT}^{\sin \phi_S }
\nonumber \\  & \quad \qquad
 +
\;  \sqrt{2\,\varepsilon (1+\varepsilon)}\, 
  \sin(2\phi_h-\phi_S)\,  
J_{2} (|\bt| |\Phperp|)\, \FTStrufu_{UT}^{\sin\lf(2\phi_h -\phi_S\rg)}
\Bigg]
\nonumber \\  &  \quad 
+
|\bm{S}_\perp| \lambda_e\, \Bigg[
  \sqrt{1-\varepsilon^2}\, \cos(\phi_h-\phi_S)\, 
J_{1} (|\bt| |\Phperp|)\, \FTStrufu_{LT}^{\cos(\phi_h -\phi_S)}
\nonumber \\  &  \quad \qquad 
+
\; \sqrt{2\,\varepsilon (1-\varepsilon)}\, 
  \cos\phi_S\, 
J_{0} (|\bt| |\Phperp|)\, \FTStrufu_{LT}^{\cos \phi_S}
\nonumber \\  &  \quad \qquad  
+
\; \sqrt{2\,\varepsilon (1-\varepsilon)}\, 
  \cos(2\phi_h-\phi_S)\,  
J_{2} (|\bt| |\Phperp|)\, \FTStrufu_{LT}^{\cos(2\phi_h - \phi_S)}
\Bigg] \Biggr\}
\label{eq:crossmaster_bspace}
\end{align}

The structure of the cross section is what one gets from a multipole expansion in $\bt$-space followed by a Fourier transform, see Appendix \ref{sec-polarFT}.
Each of the structure functions $\FTStrufu_{XY,Z}^{\cdots}$ in $\bt$-space corresponds to the Hankel (or Fourier-Bessel) transform of the corresponding structure function $F_{XY,Z}^{\cdots}$ in the usual momentum space representation of the cross section. The combinations $\sin(n\phi_h + \ldots) J_n(|\bt| |\Phperp|)$ and $\cos(n\phi_h + \ldots) J_n(|\bt| |\Phperp|)$ act as basis functions of the combined transform to $(|\Phperp|,\phi_h)$-space. Due to the fact that the multipole expansion of the physical cross section terminates, only a finite number of terms appear in the cross section, with $J_3$ being the Bessel function of highest order.
The structures $\FTStrufu_{XY,Z}^{\phantom{2}\cdots}$ are functions of $|\bt|$, $x$ and $z$, but no longer depend on the angular variables.
Introducing a short-hand notation for products
\begin{equation}
 {\cal P} [\tilde f^{(n)} \tilde D^{(m)}] \equiv \xb \sum_a e_a^2\, (zM |\bt|)^n\, (zM_h |\bt|)^m\,
\tilde f^{a(n)}(x,z^2 \bt^2) \, \tilde D^{a(m)}(z, \bt^2)\; ,
\label{eq:structure_functions}
\end{equation}
the leading twist tree level analysis in Eqs.~\eqref{eq:hadronic_tensor_bspace},~\eqref{eq:correlator_bspace} and \eqref{eq:correlator_ff_bspace}
reveals that the Fourier transformed structures in the cross section 
are simple products of \TMDPs and \TMDFs 
\begin{align}
  \FTStrufu _{UU,T} & =  
     \; {\cal P}  [ \tilde f_{1}^{(0)}  \ \tilde D_1^{(0)} ]\ , \label{eq:structurefirst}\\
 \FTStrufu_{UT,T}^{\sin(\phi_h-\phi_S)} & =    
      \; - {\cal P} [\tilde f_{1T}^{\perp (1)}  \ \tilde D_1^{(0)} ] \ , \\
\FTStrufu_{LL} & =    \; {\cal P} [\tilde g_{1L}^{(0)}  \ \tilde D_1^{(0)} ] \ ,   \\
\FTStrufu_{LT}^{\cos(\phi_h-\phi_s)} & =    \; {\cal P} [\tilde g_{1T}^{(1)}  \ \tilde D_1^{(0)} ] \ , \\
\FTStrufu_{UT}^{\sin(\phi_h+\phi_S)} & =   \; {\cal P} [\tilde h_{1}^{(0)}   \ \tilde H_1^{\perp (1)} ] \ ,\\
 \FTStrufu_{UU}^{\cos(2\phi_h)} &   = 
  \; {\cal P} [\tilde h_{1}^{\perp (1)}   \ \tilde H_1^{\perp (1)} ] \ , \\
 \FTStrufu_{UL}^{\sin(2\phi_h)} &   =    \; {\cal P} [\tilde h_{1L}^{\perp (1)}   \ \tilde H_1^{\perp (1)} ] \ ,\\
 \FTStrufu_{UT}^{\sin(3\phi_h-\phi_S)} &  =  \frac{1}{4}  {\cal P} [\tilde h_{1T}^{\perp (2)}   \ \tilde H_1^{\perp (1)} ] .\label{eq:stucturelast}
\end{align}
For completeness, we also list the above results in terms of the momentum-space structure functions ${F}_{XY,Z}^{\phantom{2}\cdots}$ of Ref. \cite{Bacchetta:2006tn} in Appendix \ref{sec-strufu}. 
Note that TMD evolution equations are typically derived in $\vprp{\elll}$-space and are thus obtained in terms of the same (derivatives of) Fourier transformed \TMDPs and \TMDFs that appear in the equations above, see, e.g., Ref. \cite{Idilbi:2004vb}, where a similar representation of the structure functions in Fourier space has been employed.

\section{Beyond tree level \label{sec-beyondtree}}
 
The formalism becomes more involved once diagrams beyond leading order in $\alpha_s$ are taken into account. 
Various strategies have been proposed to address extra divergences that appear at the one loop level and higher order~\cite{Collins:1981uk,Collins:1999dz,Collins:2003fm,Collins:2004nx,Ji:2004wu,Hautmann:2007uw,Chay:2007ty,Cherednikov:2008ua,Aybat:2011zv,CollinsBook}. The development of these frameworks for transverse momentum dependent factorization 
and the establishing of the corresponding factorization theorems is an active field of research (see e.g., Refs.~\cite{Cherednikov:2010uy, CollinsBook}). The proposed strategies require the introduction of new variables that act as regularization scales, and 
most importantly as it pertains to the content of this paper, the so called soft factors 
coming from soft-gluon radiation.
As stated in the introduction, depending on the framework, the soft factors appear explicitly in the structure functions \cite{Ji:2004xq,Ji:2004wu},  or are absorbed into the definition of \TMDPs and \TMDFs (see e.g., Refs.~\cite{CollinsBook,Aybat:2011zv}).
We will present  general arguments that soft factors cancel in weighted asymmetries, independent of the specific factorization framework; however for definiteness we work with the JMY framework~\cite{Ji:2004xq,Ji:2004wu}, which is  based on the ideas of Collins, Soper, and Sterman for the factorization of $e^+ e^- $ and Drell Yan scattering~\cite{Collins:1981uk,Collins:1984kg}.
Again we consider   the structure function 
giving rise to the Sivers asymmetry,  
\begin{align}
\FTStrufu_{UT,T}^{\sin(\phi_h-\phi_S)} & = H_{UT,T}^{\sin(\phi_h-\phi_S)}(Q^2,\mu^2,\rho) \ \tilde S^{(+)}(\bm{b}_T^2, \mu^2, \rho)\  {\cal P} [\tilde f_{1T}^{(1)} \tilde D_1^{(0)} ]  +  \tilde Y_{UT,T}^{\sin(\phi_h-\phi_S)}(Q^2,\bm{b}_T^2)\ .
 \label{eq:strucNLO}
\end{align}
The first term in the following referred to as the ``TMD expression'', 
dominates in the  region where $|\Phperp|$ is small, $ |\Phperp|/z \approx Q_T \ll Q $. The second term  is necessary to properly describe the 
structure function for 
large transverse momentum, where $Q_T\sim Q$,  and 
where fixed order perturbation theory and collinear factorization apply. 
Here $H_{UT,T}^{\sin(\phi_h-\phi_S)}$ is the hard part,  
and $\tilde S^{(+)}$ is a soft factor 
appearing explicitly in the structure function within the JMY formalism.
It is the same in all the structure functions ${\cal F}_{XY,Z}^{\phantom{2}\cdots}$, see Ref. \cite{Idilbi:2004vb}.
All other structure functions  of Eqs.\ \eqref{eq:structurefirst} - \eqref{eq:stucturelast} need to be modified analogous to Eq.~\eqref{eq:strucNLO}. 

The term $\tilde Y_{UT,T}^{\sin(\phi_h-\phi_S)}(Q^2,\bm{b}_T^2)$ represents contributions that are relevant only in the region of large transverse 
momentum $|\Phperp|$ \cite{Collins:1981va,Aybat:2011zv}.
Qualitatively, this corresponds to the very small $\vprp{b}$ 
region, $z |\bt| \lesssim 1/Q$.  Since our aim is to study \TMDPs, 
we want to focus  on the region $|\Phperp|/z \ll Q$ where  
we expect them to give 
the dominant contribution if $z|\bm{b}_T| \gg 1/Q$. 
Nevertheless, since we are considering weighted integrals of structure 
functions, the integrals do include the region of 
very large $|\Phperp|$.  
As a result, the $\tilde Y$ term in Eq.\ \eqref{eq:strucNLO} is non-zero even if $z|\bm{b}_T| \gg 1/Q$.   
We note that the $\tilde Y$ term is expected to be particularly important in 
the case of a ``mismatch'' between the tail of the TMD term and the $|\Phperp|
$-behavior obtained from the collinear formalism in the regime of intermediate 
$|\Phperp|$, i.e., $\mN \ll Q_T \ll Q$.
Matches and mismatches between the collinear and TMD formalism have been 
discussed in detail in Ref.~\cite{Bacchetta:2008xw}. An important example for 
the case of a mismatch is the $\cos(2 \phi_h)$ asymmetry.
One possibility to avoid the discussion of the $\tilde Y$-term 
is to explicitly cut off the $|\Phperp|$ integrals at some upper value
$\Lambda_\text{TMD}$. This cutoff introduces an error in our extracted TMD expression, for which we give an estimate in Appendix~\ref{sec-Ytermsuppr}.
 Another option is to simply ignore the $\tilde Y$ term. 
This amounts to keeping the TMD term in the large $|\Phperp|$
region, i.e.,\ to include the large-$|\Phperp|$-tail 
generated by the TMD term, which would otherwise be corrected by the $Y$ term.
In Appendix~\ref{sec-Ytermsuppr}, we show that in 
the $z|\bm{b}_T| \gg 1/Q$ region of  interest this produces an error 
that falls off at least as a fractional inverse power with increasing $|\bm{b}_T|$. It should be mentioned that this estimate of the behavior of 
the error applies to the Bessel weighting which we discuss below. 
By contrast, no such error estimate exists for 
conventional weighting with powers of $|\Phperp|$ since 
 such integrals are divergent.  Better error estimates, or equivalently, a better determination of the TMD region in ${\cal B}_T$, can be obtained by an explicit treatment 
of the $\tilde Y$ term, which we will leave for future analyses. 

In summary, we find that weighted integrals based on the TMD expression alone are valid only in a limited range of $\bpar$.
Finally, beyond tree level, the product notation $\mathcal{P}[f D]$ defined in  Eq.~\eqref{eq:structure_functions} has to be updated to include further dependences on the renormalization and cutoff parameters $\mu^2$, $\rho$, $\zeta$ and $\hat \zeta$  appearing in the JMY formalism 
discussed in more detail below \footnote{The framework of, e.g.,  Ref.~\cite{Aybat:2011zv}, would require analogous modifications within this formalism.}:  
\begin{align}
 {\cal P} [\tilde f^{(n)} \tilde D^{(m)}] 
 & \equiv  \xb \sum_a e_a^2 (zM |\bt|)^n (zM_h |\bt|)^m
 \tilde f^{a(n)}(x,z^2\bm{b}_T^2,\mu^2,\zeta,\rho) \tilde D^{a(m)}(z,\bm{b}_T^2,\mu^2, \hat \zeta,\rho)\; .
\end{align}

\section{\TMDPs at the level of matrix elements \label{sec-matrixel}}
 
Apart from introducing the parameters $\zeta$, $\hat \zeta$ and $\rho$
the purpose of this section is to review the formalism of Lorentz-invariant amplitudes 
underlying the decomposition of $\tilde \Phi$ Eq.\ \eqref{eq:correlator_bspace}.  
In the framework of JMY, the TMD correlator $\Phi$ itself involves a soft 
factor $S^{(+)}$ as already encountered above, i.e.,  
Eqs. \eqref{eq:phi} and \eqref{eq:phi1} need to be modified.
In the following, we  label the unmodified correlators  with the subscript  ``unsub'':  
\begin{align}
  \Phi^{[\GammaOp]}_{\text{unsub}} (\kei,P,S;v,\mu) & = \int \frac{d^4 \elll}{(2\pi)^4} \ 
  e^{i\kei \cdot \elll} 
  \underbrace{ \frac{1}{2} \bra{\nucl{P,S}}\ \bar \quark(0)\ \overbrace{ \mathcal{U}[0,\infty v]\ \mathcal{U}[\infty v,\elll] }^{\displaystyle \U} \GammaOp\ \quark(\elll)\ \ket{\nucl{P,S}} }_{\displaystyle \widetilde \Phi^{[\GammaOp]}_{\text{unsub}}(\elll,P,S;v,\mu) }\ .
  \label{eq-corrunmod}
\end{align}
The gauge link $\U$ is essentially given by two parallel straight Wilson lines running out to infinity in the direction given by the four-vector $v$ and back again. The definition of a straight Wilson line between two points $a$ and $b$ is
\begin{equation}
	\mathcal{U}[a,b] \equiv\ \mathcal{P}\ \exp\left( -ig \int_{a}^b d \xi^\mu\ A_\mu(\xi) \right) \, ,
	\label{eq-wlinecont}
\end{equation}
where $A_\mu(\xi) = T^c A_\mu^c(\xi)$, $c=1..8$ is the (matrix valued) gauge field.  A transverse link connecting these parallel Wilson lines at infinity can be omitted in the covariant gauge used by JMY.
In case of SIDIS, the direction $v=[v^-,v^+,0]$ is slightly off the light-cone direction $\nminus$, while for the Drell-Yan process $v$ is slightly off the light cone direction $-\nminus$.  The shift away from the light cone is time-like in the JMY framework and specified in a Lorentz-invariant way by the parameter $\zeta$, defined 
by $\zeta^2 = (2 P \cdot v)^2/v^2$. The parameter $\zeta$ represents a rapidity cutoff parameter \cite{Collins:1981uk}.
The above correlator can be parameterized in terms of real-valued Lorentz-invariant amplitudes. Here we restrict ourselves to the case $\Gamma=\gamma^\mu$. 
Reference \cite{Goeke:2005hb} lists the following structures
%\begin{align}
\bea
\frac{1}{2} \Phi^{[\gamma^\mu]}_{\text{unsub}} & = &P^\mu\, \Amp_2 +\kei^\mu\, \Amp_3 + \frac{1}{\mN} \epsilon^{\mu \nu \alpha \beta} P_\nu \kei_\alpha S_\beta\, \Amp_{12} + \frac{\mN^2}{(v \tcdot P)} v^\mu\, \Bmp_1 \nn
&& + \frac{\mN}{v \tcdot P} \epsilon^{\mu \nu \alpha \beta} P_\nu v_\alpha S_\beta\, \Bmp_7 + \frac{ \mN}{v \tcdot P} \epsilon^{\mu \nu \alpha \beta} \kei_\nu v_\alpha S_\beta\, \Bmp_8 \nn && + \frac{1}{\mN (v \tcdot P)} (\kei \tcdot S) \epsilon^{\mu \nu \alpha \beta} P_\nu \kei_\alpha v_\beta\, \Bmp_9  + \frac{\mN}{(v \tcdot P)^2} (v \tcdot S) \epsilon^{\mu \nu \alpha \beta} P_\nu \kei_\alpha v_\beta \Bmp_{10}\ . 
\label{eq-phidecomp}
\eea
%\end{align}
The amplitudes $\Bmp_i$ only appear when the dependence of the correlator on the direction $v$ is explicitly taken into account, and were not listed in earlier works \cite{Ralston:1979ys,Mulders:1995dh}.
Since $v$ represents only a direction, the structures above should remain invariant under re-scaling of $v$, i.e., under the substitution $v \rightarrow \eta v$, for any positive real number $\eta$. This has been ensured by dividing by powers of $v \tcdot P$ in the expression above. The amplitudes can depend on $\kei^2$, $\kei \tcdot P$, $v \tcdot \kei / (v \tcdot P)$, $v^2 / (v \tcdot P)^2 =  \zeta^{-2}$ and the sign of $v \tcdot P$, which we denote with the superscript $(+)$ (SIDIS case). For the Drell-Yan process, $v \tcdot P$ has the opposite sign $(-)$.

For our discussion below, we make use of a similar decomposition as in Eq.\ \eqref{eq-phidecomp}. However, instead of parameterizing the $\kei$-dependent correlator $\Phi^{[\GammaOp]}$, we directly parameterize the $\elll$-dependent matrix elements $\widetilde \Phi^{[\GammaOp]}$  of Eq.~\eqref{eq:phi1} in terms of complex-valued amplitudes $\tAmp_i$ and $\tBmp_i$ that depend on  $\elll^2$, $\elll \tcdot P$, $v \tcdot \elll / (v \tcdot P)$ and $\zeta^{-2}$. This parameterization in Fourier-space has already been employed in \cite{Hagler:2009mb,Musch:2010ka}
\footnote{
In Refs. \cite{Hagler:2009mb,Musch:2010ka}, a different convention for the position of the quark fields in the Fourier transformed correlator $\tilde \Phi$ has been used. These references introduce $\tilde \Phi$ as $\tilde \Phi(l,P,S,\mathcal{C}) = \frac{1}{2} \bra{\nucl{P,S}}\ \bar \quark(l)\ \mathcal{U}^\dagger[\mathcal{C}_l]\ \GammaOp\ \quark(0)\ \ket{\nucl{P,S}} $. In Eq. \eqref{eq-corrunmod} we stick to the more common convention of an operator $\overline{\quark}(0) \ldots \quark(b)$. From translation invariance follows that the variable $b$ corresponds to $-l$ in Refs.  \cite{Hagler:2009mb,Musch:2010ka}.
In particular, our amplitudes $\tilde A_i(\elll^2,\elll \tcdot P,\ldots)$ correspond to $\tilde A_i(l^2,-l \tcdot P,\ldots)$ of Refs.  \cite{Hagler:2009mb,Musch:2010ka}. 
}. 
As shown in appendix \ref{sec-amppar}, we can deduce this parameterization from Eq.\ \eqref{eq-phidecomp} using the substitution rule $\kei \rightarrow -i \mN^2 \elll$ :
\bea
  \frac{1}{2}\widetilde \Phi^{[\gamma^\mu]}_{\text{unsub}} & = & P^\mu\, \tAmp_2 - i \mN^2 \elll^\mu\, \tAmp_3 - i \mN \epsilon^{\mu \nu \alpha \beta} P_\nu \elll_\alpha S_\beta\, \tAmp_{12} + \frac{\mN^2}{(v \tcdot P)} v^\mu\, \tBmp_1\nn && + \frac{\mN}{v \tcdot P} \epsilon^{\mu \nu \alpha \beta} P_\nu v_\alpha S_\beta\, \tBmp_7  - \frac{ i \mN^3}{v \tcdot P} \epsilon^{\mu \nu \alpha \beta} \elll_\nu v_\alpha S_\beta\, \tBmp_8 \nn && - \frac{\mN^3}{v \tcdot P} (\elll \tcdot S) \epsilon^{\mu \nu \alpha \beta} P_\nu \elll_\alpha v_\beta\, \tBmp_9  - \frac{i \mN^3}{(v \tcdot P)^2} (v \tcdot S) \epsilon^{\mu \nu \alpha \beta} P_\nu \elll_\alpha v_\beta \tBmp_{10}\ . 
\label{eq-phitildedecomp}
\eea
In order to connect to the framework of \TMDPs, we integrate the correlator $\Phi$ over the (suppressed) momentum component $\kei^-$. The integration with respect to $\kei^-$ reduces the Fourier transform with respect to $\elll^+$ to the evaluation of $\widetilde{\Phi}$ at $\elll^+=0$. Moreover, in the formalism of JMY, the defining correlator of \TMDPs needs to be modified with a soft factor. The modified, $\kei^-$-integrated correlator reads
\begin{align}
  \Phi^{(+)[\GammaOp]} (x,\vprp{\kei},P,S,\mu^2,\zeta,\rho) & =
  \int \frac{d\elll^-}{(2\pi)} \ e^{ix\elll^- P^+}\ 
  \int \frac{d^2 \vprp{\elll}}{(2\pi)^2} \ e^{-i\vprp{\kei} \cdot \vprp{\elll}}\  \nonumber \\ & \times 
  \underbrace{ \frac{1}{2} \bra{\nucl{P,S}}\ \bar\psi(0)\ 
\U\ \GammaOp\ 
\quark(\elll)\ \ket{\nucl{P,S}}  }_{\displaystyle \widetilde \Phi^{[\GammaOp]}_{\text{unsub}}(\elll,P,S;v,\mu^2) }\,\Big/\, \widetilde{S}^{(+)}(\vprp{\elll}^2,\mu^2,\rho)\,\Big\vert_{\displaystyle \elll^+=0}\ ,
  \label{eq-corr}
\end{align}
where $x P^+ = \kei^+$. 
The soft factor is given as
\begin{equation}
 \widetilde{S}^{(+)}(\vprp{\elll}^2,\mu^2,\rho) = \frac{1}{N_c} \bra{0} \mathrm{Tr}_c\left\{\ \mathcal{U}[-\infty \tilde v + \elll_\perp, \elll_\perp]\ \mathcal{U}[\elll_\perp,\elll_\perp + \infty v]\ \mathcal{U}[\infty v , 0]\ \mathcal{U}[0, -\infty \tilde v]\ \right\}  \ket{0}
 \label{eq-softfdef}
\end{equation}
and involves another time-like direction $\tilde v = (\tilde{v}^-,\tilde{v}^+,0)$  slightly off the light-cone direction $\nplus$, controlled by the parameter $\rho \equiv \sqrt{v^- \tilde{v}^+/v^+ \tilde{v}^-}$. Note that $\rho^2+2+\rho^{-2}=4(v \tcdot \tilde v)^2/v^2 \tilde v^2$ is a Lorentz-invariant expression.  Here, the superscript $(+)$ specifies the sign of $v \tcdot \tilde v$, which is different for the SIDIS and the Drell-Yan process. 

In the formalism of JMY, the definition of the soft factor $\widetilde{S}^{(+)}$ above both applies to the occurrence of $\widetilde{S}^{(+)}$ in the \TMDP correlator Eq.\ \eqref{eq-corr} and in the structure function Eq.\ \eqref{eq:strucNLO}. 
In the following, we will consider the case $\GammaOp = \gamma^+$. The correlator $\Phi^{(+)[\gamma^+]}$ can be decomposed into  contributions from two distinct \TMDPs:
\begin{equation}
  \Phi^{(+)[\gamma^+]} (x,\vprp{\kei},P,S,\mu^2,\zeta,\rho)  = f_1(x,\vprp{\kei}^2;\mu^2,\zeta,\rho) - \frac{\myeps^{ij}_T\, \vect{\kei}_{T\,i}\, \vect{S}_{T\,j}}{\mN}\ f_{1T}^\perp(x,\vprp{\kei}^2;\mu^2,\zeta,\rho)\, . \label{eq-phigammaplus}
\end{equation}
Strictly speaking, $f_{1T}^\perp$ should also carry the superscript $(+)$ since it has a different sign for Drell-Yan measurements \cite{Collins:2002kn}.
We now use
\begin{align}
	\vprp{\elll}^2  & = - \elll^2\Big\vert_{\elll^+=0}\,, &  
	\elll^-  & = \frac{\elll \tcdot P}{P^+}\Big\vert_{\elll^+=0}\,, &
	R(\zeta^2) \equiv \frac{\mN^2}{v \tcdot P} \frac{v^+}{P^+} & =  1 - \sqrt{1-\frac{4 \mN^2}{\zeta^2}}\, ,
\end{align}
and insert the parameterization Eq.~\eqref{eq-phitildedecomp} into Eq.~\eqref{eq-corr}. Comparing with Eq.~\eqref{eq-phigammaplus} allows us to write the \TMDPs $f_1$ and $f_{1T}^\perp$ as
 
\begin{align}
  f_1(x,\vprp{\kei}^2;\mu^2,\zeta,\rho) = & 
%\nonumber \\ & \hspace{-3cm}
 2 \fourint \nn
& \hspace{5.5cm}\times\frac{ \tAmp_{2B}\left(\elll^2, \elll \tcdot P, \frac{(\elll \tcdot P)  R(\zeta^2)}{\mN^2}, \zeta^{-2},\mu^2\right) }{\tilde S^{(+)}(-b^2,\mu^2,\rho)}  \,    \label{eq-f1FT} 
\end{align}

\begin{align}
	f_{1T}^\perp(x,\vprp{\kei}^2;\mu^2,\zeta,\rho) & = 
4 \mN^2 \frac{\partial}{\partial (\vprp{\kei}^2)} \fourint \nn 
& \hspace{5.5cm}\times \frac{ \tAmp_{12B}\left(\elll^2, \elll \tcdot P, \frac{(\elll \tcdot P)  R(\zeta^2)}{\mN^2} , \zeta^{-2},\mu^2\right) }{\tilde S^{(+)}(-b^2,\mu^2,\rho)} 
\nonumber \\ &  = 
2 \mN^2 \int \frac{d(\elll \tcdot P)}{(2\pi)} \ e^{ix(\elll \tcdot P)}
\int_0^\infty \frac{d (-\elll^2)}{4\pi} \frac{J_1(\sqrt{-\elll^2 \vprp{\kei}^2})}{\sqrt{-\elll^2 \vprp{\kei}^2}}
  \nn
& \hspace{5.5cm} \times  \elll^2 \frac{ \tAmp_{12B}\left(\elll^2, \elll \tcdot P, \frac{(\elll \tcdot P)  R(\zeta^2)}{\mN^2} , \zeta^{-2},\mu^2\right) }{\tilde S^{(+)}(-b^2,\mu^2,\rho)} 
	 ,  \label{eq-f1TprpFT}
\end{align}
where
\begin{align}
	\tAmp_{2B} &\ \equiv \ \tAmp_2 + R(\zeta^2) \tBmp_1 \, , \\
	 \tAmp_{12B} &\ \equiv\  \tAmp_{12} - R(\zeta^2) \tBmp_8 \, .
\end{align}
We observe that the amplitudes $\tilde B_i$ give rise to structures in Eqs.\ \eqref{eq-f1FT} and \eqref{eq-f1TprpFT} that are suppressed by their explicit $\zeta$-dependence as $\zeta \rightarrow \infty$, i.e., in the limit of light-like $v$. The structures also disappear in the limit of vanishing nucleon mass $\mN^2\rightarrow0$.
Notice that the two independent Fourier transforms in each of Eqs.\ \eqref{eq-f1FT} and \eqref{eq-f1TprpFT} naturally connect the \TMDPs to a manifestly Lorentz-invariant framework and reveal $x \leftrightarrow b\tcdot P$ and $\vprp{p}^2 \leftrightarrow \elll^2$ to be pairs of conjugate variables. 

The relations of the amplitudes $\tAmp_{2B}$ and $\tAmp_{12B}$ to the $\vprp{\elll}$-Fourier-transformed \TMDPs defined in Eqs.\ \eqref{eq-bttransf} and \eqref{eq-btder} are given by
\begin{align}
	\tilde f_1^{(0)}(x,\vprp{b}^2;\mu^2,\zeta,\rho) &= \nn
& \hspace{-0.25cm} \frac{2}{\tilde S^{(+)}(\vprp{b}^2,\mu^2,\rho)} \int \frac{d(\elll \tcdot P)}{(2\pi)} \ e^{ix(\elll \tcdot P)}\ \tAmp_{2B}\left(-\vprp{b}^2, \elll \tcdot P, \frac{(\elll \tcdot P)  R(\zeta^2)}{\mN^2} , \zeta^{-2},\mu^2\right), \\
	\tilde f^{\perp(1)}_{1T}(x,\vprp{b}^2;\mu^2,\zeta,\rho) &= \nn 
&  \hspace{-0.25cm}\frac{-2}{\tilde S^{(+)}(\vprp{b}^2,\mu^2,\rho)} \int \frac{d(\elll \tcdot P)}{(2\pi)} \ e^{ix(\elll \tcdot P)}\ \tAmp_{12B}\left(-\vprp{b}^2, \elll \tcdot P, \frac{(\elll \tcdot P)  R(\zeta^2)}{\mN^2} , \zeta^{-2},\mu^2\right) .
\end{align}
We note that the soft factor would need to remain in the integrand if it were also dependent on $v \tcdot \elll /\sqrt{v^2}$, i.e., the ``angle'' between the Wilson lines and the vector $\elll$ separating the quark fields in the operator. 
The above result can also be obtained by comparison to the correlator $\tilde \Phi^{[\gamma^+]}(x,\bt)$ in Eq.\ \eqref{eq:correlator_bspace}, where
\begin{align}
	\tilde \Phi^{(+)[\Gamma]}(x,\bt,\mu^2,\zeta,\rho) \equiv \int \frac{d(\elll\tcdot P)}{(2\pi)P^+} \ e^{ix(\elll\tcdot P)}\ \frac{\widetilde \Phi^{[\GammaOp]}_{\text{unsub}}(\elll,P,S;v,\mu^2) }{ \widetilde{S}^{(+)}(-b^2,\mu^2,\rho)} \Big\vert_{\displaystyle \elll^+=0} \ .
\end{align}

\newcommand{\fourintx}{{\displaystyle \int\hspace{-1.1em}\mathcal{X}}\ }
\newcommand{\fourintxt}{{\textstyle \int\hspace{-0.9em}\mathcal{X}}\ }

\section{Bessel-weighted asymmetries \label{sec-weightedasym}}
 
As stated earlier, transverse momentum weighted 
asymmetries~\cite{Mulders:1995dh,Kotzinian:1997wt,Boer:1997nt}  
provide a means to disentangle the convolutions in the cross section in a model independent way.
Generally, the conventional weighted asymmetries are given by
%\be
%A_{XY}^{\mathcal{W}} = 2\frac{\int d |\Phperp|\, |\Phperp|\, d\phi_h\, d\phi_S\,  {\mathcal W}(|\Phperp|,\phi_h,\phi_S)
%\left(d\sigma^{XY}(\phi_h,\phi_S) \mp d\sigma^{XY}(\phi_h,\phi_S)\right)}{\int d |\Phperp|\, |\Phperp|\, d\phi_h\, d\phi_S 
%\left(d\sigma^{XY}(\phi_h,\phi_S) + d\sigma^{XY}(\phi_h,\phi_S)\right)}\;,
%\label{eq:wssa}
%\ee
\begin{align}
A_{XY}^{\mathcal{W}} & = \left\{ \begin{array}{ll}
\displaystyle
2\,\frac{\int d |\Phperp|\, |\Phperp|\, d\phi_h\, d\phi_S\,  {\mathcal W}(|\Phperp|,\phi_h)\ 
d\sigma_{XY}}{\int d |\Phperp|\, |\Phperp|\, d\phi_h\, d\phi_S\  
d\sigma_{XY}} &  \text{: for }XY=UU \\
& \\
\displaystyle
2\,\frac{\int d |\Phperp|\, |\Phperp|\, d\phi_h\, d\phi_S\,  {\mathcal W}(|\Phperp|,\phi_h,\phi_S)
\left(d\sigma_{XY}^\uparrow - d\sigma_{XY}^\downarrow\right)}{\int d |\Phperp|\, |\Phperp|\, d\phi_h\, d\phi_S 
\left(d\sigma_{XY}^\uparrow + d\sigma_{XY}^\downarrow\right)} & \text{: else,}
\end{array} 
\right.
\label{eq:wssa}
\end{align}
%\begin{align}
%A_{XY}^{\mathcal{W}} & = 2\,\frac{\left[\sum_{\lambda_e = \pm 1}(-1)^{\lambda_e}\right]_{X=L} \left[\sum_{S_\parallel = \pm 1}(-1)^{S_\parallel}\right]_{Y=L}\  \int d |\Phperp|\, |\Phperp|\, d\phi_h\, d\phi_S\,  {\mathcal W}(|\Phperp|,\phi_h,\phi_S)
%\ d\sigma_{XY}}{\left[\sum_{\lambda_e = \pm 1}\right]_{X=L} \left[\sum_{S_\parallel = \pm 1}\right]_{Y=L}\  \int d |\Phperp|\, |\Phperp|\, d\phi_h\, d\phi_S 
%\ d\sigma_{XY}}\;, 
%\label{eq:wssa} 
%\end{align}
where the labels $X,Y$ represent the polarization,  ``un'' ($U$), longitudinally ($L$) 
and transversely ($T$) of the beam and target, respectively. The angles 
$\phi_S$ and $\phi_h$ specify the directions of the hadron spin polarization and the
transverse hadron momentum respectively, relative to the lepton
scattering plane.
%The cross section  $d\sigma_{XY}$  corresponds 
%to the case with spin polarizations of the beam and  target hadron as specified by $XY$.
In case of single or double spin asymmetries $d\sigma_{XY}^\downarrow$ denotes the cross section with one of the polarizations opposite than for $d\sigma_{XY}^\uparrow$, 
such that the relevant structure function is projected out from Eq.~\eqref{eq:crossmaster_bspace}. 
We have introduced the short-hand notation $\W$ which  is a function containing various powers and $\Phperp$ as well as angular dependences of the form $\sin(m\phi_h \pm n\phi_S)$ or $\cos(m\phi_h \pm n\phi_S)$.   For the conventional 
weighted Sivers asymmetry, $\W \equiv w_1\sin(\phi_h-\phi_S)$, where
$w_1=|\Phperp|/zM$ as in Eq.~\eqref{eq:BMwtasym}.

Based on the expansion of the SIDIS cross section
  in terms of Bessel functions $J_n$
of transverse momentum and impact parameter 
in Eq.~\eqref{eq:crossmaster_bspace}, 
 we exploit the orthogonality 
to  generalize the weighting procedure.  Now the weighting is of the form
\begin{align}
A_{XY}^{\mathcal{W}}(\bpar)  & = \left\{ \begin{array}{ll}
\displaystyle
2\,\frac{\int d |\Phperp|\, |\Phperp|\, d\phi_h\, d\phi_S\,  {\mathcal W}(|\Phperp|,\phi_h;\bpar )\ 
d\sigma_{XY}}{\int d |\Phperp|\, |\Phperp|\, d\phi_h\, d\phi_S\  
J_0(|\Phperp| \bpar)\ d\sigma_{XY}} &  \hspace*{-1.5cm}\text{: for }XY=UU \\
& \\
\displaystyle
2\,\frac{\int d |\Phperp|\, |\Phperp|\, d\phi_h\, d\phi_S\,  {\mathcal W}(|\Phperp|,\phi_h,\phi_S;\bpar)
\left(d\sigma_{XY}^\uparrow - d\sigma_{XY}^\downarrow\right)}{\int d |\Phperp|\, |\Phperp|\, d\phi_h\, d\phi_S\  
J_0(|\Phperp| \bpar)\ \left(d\sigma_{XY}^\uparrow + d\sigma_{XY}^\downarrow\right)} & \text{: else,}
\end{array} 
\right.
\label{eq:wssab}
\end{align}
%\begin{align}
%A_{XY}^{\mathcal{W}}(\bpar) & = 2\,\frac{\int d |\Phperp|\, |\Phperp|\, d\phi_h\, d\phi_S\,  {\mathcal W}(|\Phperp|,\phi_h,\phi_S)\ 
%d\sigma_{XY}}{\int d |\Phperp|\, |\Phperp|\, d\phi_h\, d\phi_S\  J_0(|\Phperp| \bpar)\ 
%d\sigma_{XY}} & & \text{for }XY=UU\;, \\
%A_{XY}^{\mathcal{W}}(\bpar) & = 2\frac{\int d |\Phperp|\, |\Phperp|\, d\phi_h\, d\phi_S\,  {\mathcal W}(|\Phperp|,\phi_h,\phi_S,\bpar)
%\left(d\sigma_{XY}^\uparrow - d\sigma_{XY}^\downarrow\right)}{\int d |\Phperp|\, |\Phperp|\, d\phi_h\, d\phi_S J_0(|\Phperp| \bpar) \left(d\sigma_{XY}^\uparrow + d\sigma_{XY}^\downarrow\right)}& & \text{for } X,Y \in \{L,T\}\; ,
%\label{eq:wssab}
%\end{align}
where the weight function ${\cal W}$ corresponds to that of conventional weighted asymmetries, except that we replace
\begin{equation}
	 |\Phperp|^n \rightarrow J_n(  |\Phperp| \bpar )\, n! \left(\frac{2}{\bpar}\right)^n \; .
 \label{eq:weightfactor}
\end{equation}
As mentioned earlier, taking the asymptotic form 
of the Bessel function  the conventional weights~\cite{Kotzinian:1997wt,Boer:1997nt} which are $\propto
|\Phperp|^n$ appear as the leading term of the Taylor expansion of the right hand side of Eq.~\eqref{eq:weightfactor}.  Furthermore we note that the parameter $\bpar>0$  regularizes UV divergences in moments of \TMDPs and \FFs.
More importantly, we will show that the parameter $\bpar > 0$ allows us to scan \TMDPs and \TMDFs in Fourier space. In fact, the form of Eq.\ \eqref{eq:wssab} already indicates that the weighting implements a Fourier-decomposition of the cross section in transverse momentum space. 

Now we summarize the cancellation of the soft factor.
We will illustrate this for the Sivers Bessel-weighted asymmetry (for details see Appendix~\ref{sec-bwa}).  One can see from 
Eq.~(\ref{eq:crossmaster_bspace}) that the 
appropriate weight for the Sivers asymmetry is
\begin{equation}
\mathcal{W} = \frac{2\, J_1(|\Phperp| \bpar)}{\zh M \bpar} \sin(\phi_h-\phi_S) \text{,\quad i.e., \quad} w_1=\frac{2\, J_1(|\Phperp| \bpar)}{\zh M \bpar} \,	,
\end{equation}
corresponding to $|\Phperp|/\zh M$ in the limit $|\Phperp| \ll 1/\bpar$. Then the Bessel-weighted Sivers asymmetry is 
\begin{align}
A_{UT}^{\frac{2\, J_1(|\Phperp| \bpar)}{\zh M \bpar}\sin(\phi_h - \phi_S)}
(\bpar) & =\nn
& \hspace{-2cm}2\frac{\int d |\Phperp|\, |\Phperp|\, d\phi_h\, d\phi_S\, 
\frac{2\, J_1(|\Phperp| \bpar)}{\zh M \bpar}
 \sin(\phi_h - \phi_S) \left(d\sigma^{\uparrow} - d\sigma^{\downarrow}\right)}{\int d |\Phperp|\, |\Phperp|\, d\phi_h\, d\phi_S\ J_0(|\Phperp|\, \bpar  )\  \left(d\sigma^{\uparrow} + d\sigma^{\downarrow}\right)}\;,
\label{eq:ssa_sivers}
\end{align}
where the axially symmetric denominator is given by
\bea
\frac{2\alpha^2}{\xb y\slim Q^2}\, \frac{y^2}{(1-\varepsilon)}\,  \left( 1+\frac{\gamma^2}{2\xbj} \right) \int d |\Phperp|\, |\Phperp|\, d\phi_h\, d\phi_S\  J_0( |\Phperp|\bpar  )&& \nn && \hspace{-5cm}\times
\int \frac{d |\bt|}{(2\pi)} |\bt|  
J_{0} (|\bt| |\Phperp|)\, \FTStrufu_{UU,T} ,
\label{eq:ssa_sivers_denominator}
\eea
and from Eq.~\eqref{eq:crossmaster_bspace}  the numerator is
\bea
\frac{2\alpha^2}{\xb y\slim Q^2}\frac{y^2}{(1-\varepsilon)} 
\left( 1+\frac{\gamma^2}{2\xbj} \right)
\int d |\Phperp|\, |\Phperp| d\phi_h\, d\phi_S
\frac{2\, J_1(|\Phperp| \bpar)}{\zh M \bpar} \sin^2(\phi_h - \phi_s) &&\nn && \hspace{-7cm}\times
\int \frac{d |\bt|}{(2\pi)} |\bt|  
J_{1} (|\bt| |\Phperp|)\, \FTStrufu_{UT}^{\sin(\phi_h - \phi_s)} \, .
\label{eq:ssa_sivers_numerator}
\eea
Finally, making use of the closure relation of the Bessel function (see Appendix~\ref{appendix:derivation}) we obtain for fixed $x,y,z$, cancellation of the soft factor $S^{+(0)}(\mu^2, \rho)$ in Eq. \eqref{eq:strucNLO} from the Bessel-weighted Sivers asymmetry,
\begin{align}
 A_{UT,T}^{\frac{2\, J_1(|\Phperp| \bpar)}{\zh M \bpar}
\sin(\phi_h - \phi_s)}(\bpar)  &= \nn
&\hspace{-2.5cm} -2\frac{\sum_a e_a^2\,H_{UT,T}^{\sin(\phi_h-\phi_S)}(Q^2, \mu^2, \rho)\,  \tildeftperp(x,z^2 \bpar^2;\mu^2, \zeta,\rho)\,  \tilde D_1^{(0)a}(z,\bpar^2; \mu^2, \hat \zeta,\rho)  
}{
\sum_a e_a^2\, H_{UU,T}^{}(Q^2, \mu^2, \rho)\, \tilde f_{1}^{(0)a}(x,z^2 \bpar^2;\mu^2,\zeta,\rho)\,  \tilde D_{1}^{(0)a}(z,\bpar^2; \mu^2, \hat \zeta,\rho) 
}.
\label{eq:ssa_sivers_final2}
\end{align}
Some comments are in order.  First,  
if $|\bpar|$ is large enough, the estimate  $\tilde Y_{UT,T}^{\sin(\phi_h-\phi_S)} \sim \bpar^{-1/2}$ can be applied and may indicate that the $\tilde Y$-terms are sufficiently suppressed to be neglected for practical purposes (see  Appendix  \ref{sec-Ytermsuppr}), which is what we have done in the above equation. The above result for the Sivers asymmetry can be generalized to any other asymmetry in 
the SIDIS cross section, Eq.~\eqref{eq:crossmaster_bspace}. 
We summarize those results with the full kinematic dependences in Appendix \ref{sec-bwa}.   Weighting with Bessel functions at various values of  $\bpar$  thus allows us to map out,  ratios of Fourier-transformed \TMDPs as well as azimuthal and spin asymmetries.

Secondly, the hard scattering factor $H_{UT,T}^{\sin(\phi_h-\phi_S)}$ is expected to be the same as the unpolarized one $H_{UU,T}^{}$, 
because the Sivers effect concerns unpolarized quarks which leads to unpolarized scattering on the partonic level. This expectation is confirmed 
in a recent calculation by Kang, Xiao and Yuan~\cite{Kang:2011mr} at the one loop level, but should hold to all orders. Since this feature of the Sivers asymmetry is not shared by the other asymmetries, we will stick to writing $H_{UT,T}^{\sin(\phi_h-\phi_S)}$ to avoid potential mistakes.

Thirdly, it is important to note  that in the limit $\bpar\rightarrow0$, the cancellation of the soft factor becomes trivial, since the soft factor $\tilde{S}^+(\vprp{b},\mu,\rho)$ is unity at $\vprp{b}=0$. This has been shown in Ref.~\cite{Collins:1981uk}, but it can also be seen easily from its formal definition in terms of Wilson lines given in Eq.~\eqref{eq-softfdef}. 
Using $\tilde S^{+(0)}(\vect{0},\mu^2, \rho)=1$ shows that the $\Phperp$-integrated cross section 
does not depend on the soft factor,  as expected because the 
collinear factorization result should in principle be retrieved (after a proper regularization, which is a highly 
nontrivial matter as discussed in \cite{Collins:2003fm}).  Due to the asymptotic properties of Bessel functions for small arguments, we recover conventional weighted asymmetries in the limit $\bpar \rightarrow 0$
\bea
 A_{UT,T}^{\frac{|\Phperp|}{\zh M}\sin(\phi_h - \phi_s)}(\xbj, \zh, y) &=& \nn
&&\hspace{-3cm} 
-2 \frac{\sum_a e_a^2\ H_{UT,T}^{\sin(\phi_h-\phi_S)}(Q^2, \mu^2, \rho)\  \ftperp(x;\mu^2, \zeta,\rho) \ D_1^{a(0)}(z; \mu^2, \hat \zeta,\rho) 
}{
\sum_a e_a^2\ H_{UU,T}^{}(Q^2, \mu^2, \rho)\ f_{1}^{a(0)}(x;\mu^2, \zeta,\rho)\ D_{1}^{a(0)}(z; \mu^2, \hat \zeta,\rho) 
}\;,
\label{eq:ssa_sivers_finalbpar00}
\eea
where $ \ftperp $, $f_{1}^{a(0)}$, and $  D_{1}^{a(0)} $ are moments of \TMDPs and fragmentation functions as defined in Eq.\ \eqref{eq-btder2}.
We caution the reader that these moments are not well-defined, since the corresponding integrals are known to fall off too slowly at large transverse momentum \cite{Bacchetta:2008xw}.
Furthermore, the arguments made earlier that the $\tilde Y$-terms are small are no longer applicable.

Lastly, we briefly address what is known about the energy scale dependence of the conventional weighted asymmetries. The current knowledge on this is limited to the one-loop level. Choosing the factorization scale $\mu=Q$ 
removes the $Q$ dependence from the hard scattering function $H$ that is a function of $\ln Q^2/\mu^2$. 
This will lead to a $Q$ dependence in the transverse moments 
of the \TMDPs only~\cite{CollinsBook,Aybat:2011zv}. The scale dependence of $f_1^{(0)}(x;Q^2)$ is known, assuming that a proper definition of the \TMDP can be used, 
such that the zeroth moment corresponds to the collinear function $f_1(x; Q^2)$ after the regularization is removed. 
The same applies to $D_1^{(0)}(z;Q^2)$. For the first moment of the Sivers function one can exploit that it is directly related to 
the Qiu-Sterman function $T_F(x,x)$~\cite{Qiu:1991pp} as shown in Ref.~\cite{Boer:2003cm}. The evolution equation of the 
Qiu-Sterman function has recently been obtained \cite{Kang:2008ey,Zhou:2008mz,Vogelsang:2009pj,Braun:2009mi} 
allowing for evolution of the weighted Sivers asymmetry. The evolution of $T_F(x,x)$ is not autonomous, since it 
depends not just on $T_F(x,x)$ itself. This is true even in the large-$N_c$ limit, but in the large-$x$ limit it does 
become autonomous \cite{Braun:2009mi,Ratcliffe:2009pp}. It indicates that $f_{1T}^{\perp (1)}(x)$ evolves logarithmically with 
$Q^2$ just like $f_1(x)$, only falling off faster at a given $x$ value as $Q^2$ increases.  The evolution has also been calculated for moments of other \TMDPs such as $h_1^{\perp(1)}$~\cite{Zhou:2008fb,Zhou:2008mz,Zhou:2009jm} 
and is similar to that of $f_{1T}^{\perp (1)}$ but simpler since nonsinglet. 
In  addition, the evolution of the first moment of the Collins function,  
$H_1^{\perp(1)}$ is calculated in  \cite{Yuan:2009dw,Kang:2010xv}. 

\section{Average transverse momentum shift and Bessel-weighted counterpart \label{sec-avgtransmom}}

In a similar manner to Section \ref{sec-weightedasym} we now consider the soft factor cancellation in the average transverse momentum shift of unpolarized quarks in a transversely polarized nucleon for a given longitudinal momentum fraction $x$. 
This shift is considered in \cite{Burkardt:2003uw} and  defined 
by a ratio of the $\vprp{\kei}$-weighted correlator:
\begin{align}
	\langle \kei_y(x) \rangle_{TU} &  = \left. \frac{ \int d^2 \vprp{\kei}\, \vect{\kei}_y  \ \Phi^{(+)[\gamma^+]}(x,\vprp{\kei},P,S,\mu^2,\zeta,\rho) }{ \int d^2 \vprp{\kei}  \phantom{\vect{\kei}_y}\ \Phi^{(+)[\gamma^+]}(x,\vprp{\kei},P,S,\mu^2,\zeta,\rho) } \right|_{S^\pm=0,\,\vprp{S}=(1,0)} 
	= \mN \frac{ f_{1T}^{\perp(1)}(x;\mu^2,\zeta,\rho) }{ f_1^{(0)}(x;\mu^2,\zeta,\rho)} \, , 
	\label{eq:ktmomratio}
\end{align}
where $ f_{1T}^{\perp(1)}$ and $f_1^{(0)}$ are the moments defined in Eqs.\ \eqref{eq-btder2}. Obviously, the average momentum shift is very similar in structure to the weighted asymmetry Eq.\ \eqref{eq:ssa_sivers_finalbpar00}. While the weighted asymmetries are accessible directly from the $\Phperp$-weighted cross section, the average transverse momentum shifts are obtained from the $\vprp{\kei}$-weighted correlator and could in principle be accessible from weighted jet asymmetries. As already mentioned, the integrals defining the moments of \TMDPs on the right hand side of the above equation are divergent without suitable regularization. In the following, we therefore generalize the above quantity, weighting with Bessel functions of $|\vprp{\kei}|$ instead. In particular, we replace
\begin{equation}
	\vect{\kei}_y = |\vprp{\kei}| \, \sin(\phi_\kei) \quad \longrightarrow \quad 
  \frac{2 J_1( |\vprp{\kei}| \bpar )}{\bpar} \, \sin(\phi_\kei-\phi_S ) \, ,
\end{equation}
where $\phi_S=0$  for the choice $\vprp{S}=(1,0)$ in Eq. \eqref{eq:ktmomratio}. The correlator $\Phi^{(+)[\gamma^+]}$ reads in terms of the amplitudes $\tAmp_i$
and $\Bmp_i$, 
\begin{align}
	\Phi^{(+)[\gamma^+]}(x,\vprp{\kei},P,S,\mu^2,\zeta,\rho) & = \fourintx \int_0^\infty \frac{d |\vprp{\elll}|}{2\pi} |\vprp{\elll}| \ \Big\{\  
		J_0(|\vprp{\elll}|\, |\vprp{\kei}|)\, 2\tAmp_{2B}/\widetilde{S} \nonumber \\
		& - M  |\vprp{\elll}|\, |\vprp{S}|\, \sin(\phi_\kei-\phi_S)\, J_1(|\vprp{\elll}|\, |\vprp{\kei}|)\,  2\tAmp_{12B}/\widetilde{S} 
	\ \Big\}\, ,
\end{align}
where we abbreviate
\begin{equation}
	\fourintx \equiv \int \frac{d(\elll \tcdot P)}{(2\pi)} \ e^{ix(\elll \tcdot P)}\, .
\end{equation}
The Bessel-weighted analog of Eq.\ \eqref{eq:ktmomratio} is thus
\begin{align}
	\langle \kei_y(x) \rangle_{TU}^{\bpar}
	& \equiv\ \left. \frac{
	  \int d|\vprp{\kei}|\, |\vprp{\kei}| \int d \phi_\kei
\frac{2\, J_1( | \vprp{\kei}| \bpar )}{\bpar}
\sin(\phi_\kei - \phi_S)\ \Phi^{(+)[\gamma^+]}(x,\vprp{\kei},P,S,\mu^2,\zeta,\rho) }
	{ \int d|\vprp{\kei}|\, |\vprp{\kei}| \int d \phi_\kei
J_0( | \vprp{\kei}| \bpar )\, \phantom{\sin(\phi_\kei - \phi_S) } \Phi^{(+)[\gamma^+]}(x,\vprp{\kei},P,S,\mu^2,\zeta,\rho) } \right|_{|\vprp{S}|=1} \nonumber \\
	& =- M \frac{\fourintx \tAmp_{12B}\left(-\bpar^2,\elll \tcdot P, \frac{(\elll \tcdot P)  R(\zeta^2)}{\mN^2} , \zeta^{-2},\mu^2\right)}
		{\fourintx \tAmp_{2B}\left(-\bpar^2,\elll \tcdot P, \frac{(\elll \tcdot P)  R(\zeta^2)}{\mN^2} , \zeta^{-2},\mu^2\right)}\nonumber \\
	& = \mN \frac{ \tilde f_{1T}^{\perp(1)}(x,\bpar^2;\mu^2,\zeta,\rho) }{ \tilde f_1^{(0)}(x,\bpar^2;\mu^2,\zeta,\rho)} \, .
	\label{eq:ktmomratio_Bessel}	
\end{align}
Again, the soft factors cancel. At this point, the independence of the soft factor on $v \tcdot \elll / \sqrt{v^2}$ is crucial.
In the limit $\bpar \rightarrow 0$, we recover equation \eqref{eq:ktmomratio}, $\langle \kei_y(x) \rangle_{TU}^0 =  \langle \kei_y(x) \rangle_{TU}$ , which we have thus shown to be formally free of any soft factor contribution. However, we caution the reader again that the expressions at $\bpar = 0$ can be ill-defined without an additional regularization step. 

We can go one step further and form ratios that are also integrated in $x$, with weights $\exp(-i x \bparx)$.
For $\bparx = 0$, this is the same as taking the lowest $x$-moment 
that appears in the Burkardt sum rule \cite{Burkardt:2003uw}. 
The reason  it is interesting to look at such quantities is their renormalization properties. 
Another motivation to discuss such quantities here is lattice QCD. Taking $x$-moments is a standard ingredient in lattice computations of nucleon structure, see e.g., Ref. \cite{Hagler:2009ni} for a review. First exploratory studies of \TMDPs on the lattice \cite{Hagler:2009mb,Musch:2010ka} focus to a large degree on computations of the lowest $x$-moment of distributions, but access to finite values of $\bparx$ is also possible.
By ``integration over $x$'' we mean an integration over the entire support of the correlator; this includes contributions from negative $x$ which correspond to anti-quark contributions, see e.g., Ref. \cite{Mulders:1995dh,Musch:2010ka} for details.
In particular, the $x$-integrals of the two \TMDPs $f_1$ and $f_{1T}^\perp$ can be decomposed as
\begin{align}
	 \int_{-1}^1 dx\ e^{-i x \bparx}\  f_1(x,\vprp{\kei}^2;\mu^2,\zeta,\rho) 
& =  \nn &\hspace{-1cm} 
\int_0^1 dx\  \left\{ e^{-i x \bparx} f_1(x,\vprp{\kei}^2;\mu^2,\zeta,\rho)  -  e^{i x \bparx} \bar{f}_1(x,\vprp{\kei}^2;\mu^2,\zeta,\rho) \right\} \, ,\nonumber \\
	 \int_{-1}^1 dx\ e^{-i x \bparx}\ f_{1T}^{\perp}(x,\vprp{\kei}^2;\mu^2,\zeta,\rho) 
& =  \nn &\hspace{-1cm} 
 \int_0^1 dx\ \left\{  e^{-i x \bparx} f_{1T}^{\perp}(x,\vprp{\kei}^2;\mu^2,\zeta,\rho) +  e^{i x \bparx} \bar{f}_{1T}^{\perp}(x,\vprp{\kei}^2;\mu^2,\zeta,\rho) \right\} , 	 
\end{align}
where $\bar{f}_1$ and $\bar{f}_{1T}^\perp$ are anti-quark distributions.
In analogy to Eq.\ \eqref{eq:ktmomratio_Bessel}, one can consider
\bea
	\langle \kei_y \rangle_{TU}^{\bparx,\bpar}  
%	& \equiv &\nn &&
%\hspace{-1cm} 
&\equiv &\frac{
	  \int d|\vprp{\kei}| |\vprp{\kei}| \int d \phi_\kei\ \frac{2J_1( | \vprp{\kei}| \bpar )}{\bpar} \sin(\phi_\kei -\phi_S)
\int dx e^{-i x \bparx}\Phi^{\scriptscriptstyle(+)[\gamma^+]}{\scriptstyle{(x,\vprp{\kei},P,S,\mu^2,\zeta,\rho)}}}{\int d|\vprp{\kei}| |\vprp{\kei}| \int d \phi_\kei\ J_0( | \vprp{\kei}|\bpar ) \phantom{\sin(\phi_\kei - \phi_S)}%}
\int dx e^{-i x \bparx} \Phi^{\scriptscriptstyle(+)[\gamma^+]}{\scriptstyle{(x,\vprp{\kei},P,S,\mu^2,\zeta,\rho)}}}
%\nn && \hspace{4.5cm}
%\frac{
%\Bigg|_{|\vprp{S}|=1} 
\nonumber \\
	& = &\mN \frac{ \int dx e^{-i x \bparx}\ \tilde f_{1T}^{\perp(1)}(x,\bpar^2;\mu^2,\zeta,\rho) }{ \int dx e^{-i x \bparx}\ \tilde f_1^{(0)}(x,\bpar^2;\mu^2,\zeta,\rho) } \nonumber \\
	& = & -M \frac{\tAmp_{12B}\left(-\bpar^2, \bparx, \bparx \frac{R(\zeta^2)}{\mN^2}, \zeta^{-2}, \mu^2 \right)}
		{\tAmp_{2B}\left(-\bpar^2,\bparx , \bparx \frac{R(\zeta^2)}{\mN^2}, \zeta^{-2},\mu^2\right)}\, .
	\label{eq:ktmomratio_Bessel_x}	
\eea
In this case, the cancellation of the soft factor occurs even for a soft factor that has a dependence on $v \tcdot \elll / \sqrt{v^2}$.
In the last line of Eq.\ \eqref{eq:ktmomratio_Bessel_x}, the amplitudes in the numerator and denominator parameterize the same matrix element $\tilde \Phi^{[\gamma^+]}(\elll,P,S;v,\mu)$, 
hence they involve the same bi-local quark-quark operator.
The work on renormalization properties of non-local operators involving Wilson lines in Refs. \cite{Dotsenko:1979wb,Craigie:1980qs,Arefeva:1980zd,Aoyama:1981ev,Stefanis:1983ke,Dorn:1986dt} suggests that the operator $\overline{\quark}(0) \U \quark(\elll)$ might renormalize multiplicatively for not too small $|\vprp{\elll}|$, compare also \cite{Cherednikov:2008ua}. 
As a result, the quantity $\langle \kei_y \rangle_{TU}^{\bpar,\bparx}$ would be renormalization scheme and scale independent (up to the evolution with the rapidity cutoff parameter $\zeta^2$), since all multiplicative renormalization factors would cancel in the ratio. This observation was already made in Ref.\ \cite{Musch:2010ka} and is consistent with TMD factorization, which also involves only multiplicative renormalization for 
$|\Phperp| \ll Q$ or $\bpar \gg 1/Q$. For smaller $\bpar$, mixing with gluonic operators is expected, as it is known that the Qiu-Sterman function for quarks and gluons (related to $\tilde f_{1T}^{\perp (1) q,g}(x,\vprp{\elll}^2=0)$) mix under changes of the scale \cite{Kang:2008ey,Zhou:2008mz,Vogelsang:2009pj,Braun:2009mi}, thereby preventing the cancellation of multiplicative factors in the ratio
considered here. The properties of quantities like $\langle \kei_y \rangle_{TU}^{\bpar,\bparx}$ remain to be studied more thoroughly. They could be interesting objects to make contact between theory predictions from, e.g., lattice QCD, and experiment.

\section{Conclusions \label{sec-conl}}

We have shown that rewriting the SIDIS cross-section in coordinate space 
displays the important feature that structure functions become simple products of  
Fourier transformed \TMDPs\ and FFs, or derivatives thereof.  The angular structure of the cross section naturally suggests weighting with Bessel functions in order to project out these 
Fourier-Bessel transformed distributions, which serve as well-defined replacements of the transverse moments
entering conventional weighted asymmetries. In addition, Bessel-weighted asymmetries 
provide a unique opportunity to study nucleon structure in a model independent way due 
to the absence of the soft factor $S^{+(0)}$, which as we have shown cancels from these observables. 
This cancellation is based on the fact that the soft factor 
is flavor blind in hard processes, and it depends only on $\bm{\elll}^2_T, \mu^2, \rho$. 
Moreover, evolution equations for the distributions are typically calculated in terms of the (derivatives of) Fourier transformed \TMDPs and FFs. As a result the study of the scale dependence of Bessel-weighted asymmetries should prove more straightforward.
For the above stated reasons we propose Bessel-weighted asymmetries as clean observables to study the scale 
dependence of \TMDPs and FFs at existing (HERMES, COMPASS, JLab) and future facilities (Electron Ion Collider, JLab 12 GeV).
Our results are also easily generalized to other processes where TMD factorization is valid, such as 
$e^+e^-$ annihilation and Drell-Yan processes. 

\section{Acknowledgments}
We thank Elke Aschenauer, Harut Avakian,
Mert Aybat, Vladimir Braun, Matthias Burkardt, Maarten Buffing, John Collins,
Markus Diehl, Rolf Ent, Philipp H\"agler, Aram Kotzinian,
Andreas Metz, Piet Mulders,   Ted Rogers,  and Feng Yuan for fruitful discussions.
We are grateful for partial support from the Institute For Nuclear Theory (INT), University of Washington and to the organizers of the INT Workshop, "Gluons and the Quark Sea at High Energies: Distributions, Polarization, Tomography" where part of this work was undertaken. LG acknowledges support from  U.S. Department of Energy under contract DE-FG02-07ER41460, and thanks the JLab theory group for support.  Authored by Jefferson Science Associates, LLC under U.S. DOE Contract No. DE-AC05-06OR23177. 
The U.S. Government retains a non-exclusive, paid-up, irrevocable, world-wide license to publish or reproduce this manuscript for U.S. Government purposes.

%%%%%%%%%%%%%%%%%%%%%%%%%%%%%%%% APPENDIX %%%%%%%%%%%%%%%%%%%%%%%%%%%%%%%%

\appendix

\section{Conventions and useful relations}
\label{sec-convent}

In Section~\ref{sec-crosssec}
we use definitions for the kinematic variables and the ratio of  longitudinal and transverse photon flux $\epsilon$ as in Ref. \cite{Bacchetta:2006tn},
\be
\xbj = \frac{Q^2}{2\,P\cdott q}, \;\;
y = \frac{P \cdott q}{P \cdott l}, \;\;
z_h = \frac{P \cdott P_h}{P\cdott q}, \;\;
\gamma = \frac{2 M x}{Q}, \;\; \varepsilon = \frac{1-y -\frac{1}{4}\slim \gamma^2 y^2}{1-y
  +\frac{1}{2}\slim y^2 +\frac{1}{4}\slim \gamma^2 y^2} \ ,
  \label{eq:xyz_app}
\ee
where $M$ is the mass of the target nucleon. 
The off-collinearity of the process is characterized by the variable $Q_T$ introduced through
\begin{align}
	q_T & \equiv q + (1+q_T^2/Q^2)xP - P_h/z, &
	Q_T & \equiv \sqrt{-q_T^2}
\end{align}
and for $Q_T \ll Q$, one finds $|\Phperp| \approx z Q_T$, see, e.g. \cite{Bacchetta:2008xw}.  
The leptonic tensor is 
\begin{equation}
 L_{\mu\nu} = 2 (l_\mu l'_\nu + l_\nu l'_\mu - (l\cdot l') g_{\mu\nu} + i \lambda_e \epsilon_{\mu\nu\alpha\beta}l^\alpha q^\beta)\; ,
 \end{equation}
where we neglected the lepton mass.
The hadronic tensor is  
\begin{align}
	2 M W^{\mu\nu} \hspace*{-0.1cm}=\hspace*{-0.1cm} \sum_X \hspace*{-0.1cm}\int\hspace*{-0.1cm} \frac{d^3 \vect{P}_X}{(2\pi)^3 2 P_X^0
} \delta^{(4)}\left(q{+}P{-}P_X{-}P_h\right)\bra{P,S} J_\mu(0) \ket{P_X,P_h}\bra{P_X,P_h}J_\nu(0)\ket{P,S}.
\end{align}
For an arbitrary four-vector $\omega$, we introduce the usual light-cone decomposition as
\begin{equation}
 \omega^\mu = \omega^+ \nplus^\mu + \omega^- \nminus^\mu + \omega_T^\mu\; ,
\end{equation}
where $\omega^\pm = (\omega^0\pm \omega^3)/\sqrt{2}$ and where the basis vectors $\nplus^\mu$ and $\nminus^\mu$ are
\begin{equation}
   \nplus^\mu = \frac{1}{\sqrt{2}}\left(1,0,0,1\right), \; \,  \nminus^\mu = \frac{1}{\sqrt{2}}\left(1,0,0,-1\right) ,
\end{equation}
such that $n_\pm \cdot n_\mp = 1$,  $n_\pm \cdot n_\pm = 0$, $n_\pm \cdot \omega_T = 0$. 
Note that $\omega_T\cdot \omega_T= - \vect{\omega}_T^2$.
In the  $\gamma^* P$ center of mass frame with the proton three-momentum pointing in positive $z$-direction, 
we can decompose the proton and parton momenta as
\begin{eqnarray}
P^\mu &=& P^+ n_{+}^\mu + \frac{M^2}{2 P^+} \nminus^\mu \; ,  \nonumber \\ 
p^\mu &=& x P^+ \nplus^\mu + \frac{p^2+\pt^2}{2xP^+} \nminus^\mu + {p}_T^\mu \; , \label{kin1}  
\end{eqnarray}
where  $x=p^+/P^+$ is the quark light-cone momentum fraction. 

Finally, using Eq.~\eqref{eq:bspace_trace}  in  Section~\ref{sec:FT}, 
we write the quark-quark correlator  reconstructed from the trace projections
\begin{align}
    \tilde \Phi &= \frac{1}{2} \gamma_+  \tilde \Phi^{[\gamma^+]} \,-\,  \frac{1}{2} \gamma_+ \gamma^5  \tilde \Phi^{[\gamma^+ \gamma^5]}
\,-\,  \frac{1}{4} i\sigma_{\alpha +} \gamma^5  \tilde \Phi^{[i\sigma^{\alpha +} \gamma^5]}   \,+\, \frac{1}{2} \gamma_\beta  \tilde \Phi^{[\gamma^\beta]}
\nonumber \\  &
-\,  \frac{1}{2} \gamma_\beta \gamma^5  \tilde \Phi^{[\gamma^\beta \gamma^5]} \,-\,  \frac{1}{4} i\sigma_{\alpha \beta} \gamma^5  \tilde \Phi^{[i\sigma^{\alpha \beta} \gamma^5]} \,+\, \frac{1}{2} {\Eins}  \tilde \Phi^{[\Eins]} \; ,
\label{eq-untracephi}
\end{align}
where $\alpha = 1,2$ and  $\beta=-,1,2$. Here we have also included twist-3 and twist-4 terms.
For the fragmentation correlator we have the following expression,
\begin{align}
    \tilde \Delta &=  \frac{1}{2} \gamma_-  \tilde \Delta^{[\gamma^-]} \,-\,  \frac{1}{2} \gamma_- \gamma^5  \tilde \Delta^{[\gamma^- \gamma^5]}
\,-\,  \frac{1}{4} i\sigma_{\alpha -} \gamma^5  \tilde \Delta^{[i\sigma^{\alpha -} \gamma^5]}   \,+\, \frac{1}{2} \gamma_\beta  \tilde \Delta^{[\gamma^\beta]}
\nonumber \\  & 
-\,  \frac{1}{2} \gamma_\beta \gamma^5  \tilde \Delta^{[\gamma^\beta \gamma^5]} \,-\,  \frac{1}{4} i\sigma_{\alpha \beta} \gamma^5  \tilde \Delta^{[i\sigma^{\alpha \beta} \gamma^5]} \,+\, \frac{1}{2} {\Eins} \Delta^{[\Eins]} \; .
\label{eq-untracedelta}
\end{align}

\section{Multipole expansion and Fourier transform\label{sec-polarFT}}

This appendix shows the simple underlying mathematical structure of Eq.\ \eqref{eq:crossmaster_bspace}. 
Let us treat all kinematic variables except for $\Phperp$ and $\phi_h$ as constants.
Consider the cross section $\tilde \sigma$ in Fourier space as some arbitrary function that depends on $\bt$. 
This dependence can be formulated in coordinate space $(|\bt|,\phi_b)$, 
\begin{equation}
	\tilde \sigma(|\bt|,\phi_b) = \sum_{n=-\infty}^{\infty} e^{i n \phi_b} \tilde \sigma_n(|\bt|)
\end{equation}
which is nothing but a multipole expansion with $|\bt|$ dependent coefficients $\tilde \sigma_n$.
Performing a Fourier-transform of $\tilde \sigma$ back to momentum space $(|\Phperp|,\phi_h)$, we obtain
\begin{align}
  \sigma(|\Phperp|,\phi_h) & = 
  \int \frac{d^2 \vprp{\elll}}{(2\pi)^2}\  e^{-i \Phperp\cdot \vprp{\elll}}\ \tilde \sigma(\vprp{\elll}) \nonumber \\
  & = \int \frac{d |\vprp{\elll}|}{2\pi}\, |\vprp{\elll}| \int_0^{2\pi} \frac{d \phi_\elll}{2\pi}\  e^{-i |\Phperp| |\vprp{\elll}| \cos(\phi_h - \phi_\elll) }\ \sum_{n=-\infty}^\infty e^{i n \phi_\elll}  \tilde \sigma_n(|\vprp{\elll}|) \nonumber \\
  & = \sum_{n=-\infty}^\infty  e^{i n \phi_h} \int \frac{d |\vprp{\elll}|}{2\pi} \, |\vprp{\elll}|\ 
  (-i)^n  J_n(|\Phperp| |\vprp{\elll}|)\  \tilde \sigma_n(|\vprp{\elll}|) \, .
  \end{align}
Using $e^{in\phi_h} = \cos(n\phi_h)+i\sin(n\phi_h)$ and $J_n = (-1)^n J_{-n}$, it is evident that the last line of the above equation has exactly the form of the cross section Eq.\ \eqref{eq:crossmaster_bspace}, where a finite number of the $\tilde \sigma_{\pm n}$ is given by simple linear combinations of the structure functions $\FTStrufu_{XY,Z}^{\cdots}$, and the rest is zero.
In our case, the Bessel function $J_n$ with the highest $n$ is $J_3$, which appears in combination with the angular $\sin(3\phi_h - \phi_S)$ modulation and turns out to be associated with the quadrupole deformation of parton densities $h_{1T}^\perp$.

\section{Parameterization of the correlator in $\elll$-space \label{sec-amppar}}

First, we briefly review the relevant properties of the correlator under symmetry transformations.  Applying Lorentz transformations ($L$), parity transformation ($P$), time-reversal ($T$) and hermitian conjugation ($\dagger$) to the matrix elements, we find that the correlator fulfills
\begin{align}
	&(L): &\Phi^{[\GammaOp]}_{\text{unsub}}(\kei,P,S;v,\mu)  
	&= \Phi^{[\Lambda_{\slfrac{1}{2}}^{-1}\GammaOp\Lambda_{\slfrac{1}{2}}^{\phantom{-1}}]}_{\text{unsub}}(\Lambda \kei,\Lambda P, \Lambda S;\Lambda v,\mu)  \ ,\label{eq-lortrans} \displaybreak[0] \\
	&(P): &\Phi^{[\GammaOp]}_{\text{unsub}}(\kei,P,S;v,\mu)  
	&= \Phi^{[\gamma^0\GammaOp\gamma^0]}_{\text{unsub}}(\overline{\kei},\overline{P},-\overline{S};\overline{v},\mu)  \ ,\label{eq-conpar} \displaybreak[0] \\
	&(T): &\left[ \Phi^{[\GammaOp]}_{\text{unsub}}(\kei,P,S;v,\mu)  \right]^* 
	&= \Phi^{[\gamma^1 \gamma^3 \GammaOp^* \gamma^3 \gamma^1]}_{\text{unsub}}(\overline{\kei},\overline{P},\overline{S};-\overline{v},\mu)  \ , \label{eq-contime} \displaybreak[0] \\
	&(\dagger): &\left[  \Phi^{[\GammaOp]}_{\text{unsub}}(\kei,P,S;v,\mu)  \right]^* 
	&= \Phi^{[\gamma^0 \GammaOp^\dagger \gamma^0]}_{\text{unsub}}(\kei,P,S;v,\mu)  \ .\label{eq-conherm}
	\end{align}
where we denote the sign change of spatial components of a given vector $c$; that is,  $\overline{c} \equiv (c^0, -c^1, -c^2, -c^3)$. From hermiticity $(\dagger)$ follows that the $\Amp_i$ and $\Bmp_i$ in Eq.\ \eqref{eq-phidecomp} are real valued. Time reversal $(T)$ does not constrain the number of allowed structures, because it changes the sign of $v\tcdot P$. Instead, time reversal $(T)$ establishes relations between SIDIS amplitudes $A_i^{(+)}$, $B_i^{(+)}$ and Drell-Yan amplitudes $A_i^{(-)}$, $B_i^{(-)}$.
	
For any of the transformations $\mathcal{T} \in \{ L, P, T, \dagger \}$, the Eqs.\ \eqref{eq-lortrans}-\eqref{eq-conherm} are of the general form 
\begin{equation}
	\mathcal{T}_\Phi\left(\Phi(\kei,w)\right) = \Phi\left(\mathcal{T}_\kei(\kei),\mathcal{T}_w(w)\right)
\end{equation}
where we have omitted the subscript ``unsub'' and the renormalization scale $\mu$, and where the symbol $w$ summarizes all dependences on $\GammaOp$, $P$, $S$ and $v$. Here $\mathcal{T}_\Phi$ is either the identity function or complex conjugation.  The transformation rule $\mathcal{T}_\kei(\kei)$ maps onto $\Lambda \kei$, $\kei$ or $\overline{\kei}$ and thus fulfills $a \tcdot b = \mathcal{T}_\kei(a) \tcdot \mathcal{T}_\kei(b)$ for any two vectors $a$ and $b$. The Fourier-transformed correlator 
\begin{equation}
	\widetilde{\Phi}(\elll,w) =  \int d^4 \kei \ e^{-i \kei \tcdot \elll} \Phi(\kei, w)
\end{equation}
transforms according to
\begin{align}
	\mathcal{T}_\Phi\left(\widetilde{\Phi}(\elll,w) \right) & =  \int d^4 \kei \ e^{\mathcal{T}_\Phi(-i)\, \kei \tcdot \elll}\ \mathcal{T}_\Phi \left( \Phi(\kei, w) \right) \nonumber \\
	& =  \int d^4 q \ e^{\mathcal{T}_\Phi(-i)\, \mathcal{T}^{-1}_\kei(q) \tcdot \elll} \ \Phi\left(q, \mathcal{T}_w( w) \right)  \nonumber \\
	& =  \int d^4 q \ e^{\mathcal{T}_\Phi(-i)\, q \tcdot \mathcal{T}_\kei(\elll)} \ \Phi\left(q, \mathcal{T}_w( w) \right) \nonumber \\
	& = \widetilde{\Phi}\left( \frac{\mathcal{T}_\Phi(i)}{i} \mathcal{T}_\kei(\elll), \mathcal{T}_w(w) \right)\, .
	\label{eq-phitildetrafo}
\end{align}
For example, $\tilde \Phi$ transforms under hermitian conjugation as
\begin{align}
	&(\dagger): &\left[ \widetilde{\Phi}^{[\GammaOp]}_{\text{unsub}}(\elll,P,S;v)  \right]^* 
	&= \widetilde{\Phi}^{[\gamma^0 \GammaOp^\dagger \gamma^0]}_{\text{unsub}}(-\elll,P,S;v)  \ .\label{eq-conhermtilde}
\end{align}
Let $f(\kei,w)$ be any of the structures preceding the invariant amplitudes in the parameterization of $\Phi$.
The structure $f(\kei,w)$ is a homogeneous function of some degree $n$ in $\kei$, i.e., $f(\alpha \kei,w) = \alpha^n f(\kei,w)$ for any number $\alpha$. 
For example, the structure $f(\kei,w) = \frac{1}{\mN (v \tcdot P)} (\kei \tcdot S) \epsilon^{\mu \nu \alpha \beta} P_\nu \kei_\alpha v_\beta$ preceding $\Bmp_9$ in Eq.\ \eqref{eq-phidecomp} has degree $n=2$.
If we define $\tilde f(\elll,w) \equiv f(- i \mN^2 \elll,w)$, then 
\begin{align}
	\mathcal{T}_\Phi\left( \tilde f(\elll, w) \right) =
	\mathcal{T}_\Phi(-i \mN^2)^n\, \mathcal{T}_\Phi\left( f( \elll, w) \right) =
	f\left( \mathcal{T}_\Phi(-i \mN^2) \mathcal{T}_\kei(\elll) , \mathcal{T}_w(w) \right) =
	\tilde f \left(  \frac{\mathcal{T}_\Phi(i)}{i} \elll, w \right)\, .
	\end{align}
This shows that $\tilde f$ transforms like $\widetilde \Phi$ in Eq.\ \eqref{eq-phitildetrafo}.
We conclude that the parameterization of $\widetilde{\Phi}$ can be found by the substitution $\kei \rightarrow -i \mN^2 \elll$ in the structures parameterizing $\Phi$, and we arrive at Eq.\ \eqref{eq-phitildedecomp}. 
The amplitudes $\tAmp_i$ and $\tBmp_i$ introduced this way are no longer constrained to be real valued functions. Instead, hermitian conjugation Eq.\ \eqref{eq-conhermtilde} yields the relation
\begin{equation}
	\left[ \tAmp_i(\elll^2,\elll \tcdot P, v \tcdot \elll / (v \tcdot P), \zeta^{-2},\mu^2) \right]^* = \tAmp_i(\elll^2,-\elll \tcdot P, -v \tcdot \elll / (v \tcdot P), \zeta^{-2},\mu^2)\, .
\end{equation}

\section{Structure functions in terms of Fourier transformed \TMDPs and FFs \label{sec-strufu}}

The structure functions of Ref. \cite{Bacchetta:2006tn} can be expressed in terms of Fourier-transformed \TMDPs and FFs as
\begin{align}  
  F_{UU,T} & =  
 \xbj \sum_a e_a^2 \int \frac{d |\bm{b}_T|}{(2\pi)} |\bm{b}_T|\, J_0(|\bm{b}_T|\,|\Phperp| )\ \tilde f_{1}^{a}(x,z^2\bm{b}_T^2) \ \tilde D_1^{a}(z,\bm{b}_T^2)\ ,\displaybreak[0] \\
	 F_{UT,T}^{\sin(\phi_h-\phi_S)} & =   
 - \xbj \sum_a e_a^2 \int \frac{d |\bm{b}_T|}{(2\pi)} |\bm{b}_T|^2\, J_1(|\bm{b}_T|\,|\Phperp| )\ \mN z\ \tilde f_{1T}^{\perp a (1)}(x,z^2\bm{b}_T^2) \ \tilde D_1^a(z,\bm{b}_T^2)\ ,   \displaybreak[0]\\
F_{LL} & =   \xbj \sum_a e^2_a \int \frac{d |\bm{b}_T|}{(2\pi)} |\bm{b}_T|\, J_0(|\bm{b}_T|\,|\Phperp| )\ \tilde g_{1L}^a(x,z^2\bm{b}_T^2) \ \tilde D_1^a(z,\bm{b}_T^2)\ ,\label{eq:c1}\displaybreak[0] \\
F_{LT}^{\cos(\phi_h-\phi_s)} & =  \xbj \sum_a e_a^2 
\int \frac{d |\bm{b}_T|}{(2\pi)} |\bm{b}_T|^2\, J_1(|\bm{b}_T|\,|\Phperp| )\ \mN z\
 \ \tilde g_{1T}^{\perp a (1)}(x,z^2\bm{b}_T^2) \ \tilde D_1^a(z,\bm{b}_T^2) \ ,
\label{eq:c2}\displaybreak[0]\\
F_{UT}^{\sin(\phi_h+\phi_S)} & =\xbj \sum_a e_a^2 \int \frac{d |\bm{b}_T|}{(2\pi)} |\bm{b}_T|^2\, J_1(|\bm{b}_T|\,|\Phperp| )\ \mN_h z\ \tilde h_{1}^a(x,z^2\bm{b}_T^2) \ \tilde H_1^{\perp a (1)}(z,\bm{b}_T^2)\ ,
\label{eq:c3}\displaybreak[0]\\
 F_{UU}^{\cos(2\phi_h)} &   = 
\xbj  \sum_a e_a^2 \int \frac{d |\bm{b}_T|}{(2\pi)} |\bm{b}_T|^3\, 
J_2(|\bm{b}_T|\,|\Phperp| ) \mN M_h z^2\ \tilde h_{1}^{\perp a (1)}(x,z^2\bm{b}_T^2) \ \tilde H_1^{\perp a (1)}(z,\bm{b}_T^2)\ , 
\label{eq:c4}\displaybreak[0]\\
 F_{UL}^{\sin(2\phi_h)} &   = \xbj  \sum_a e_a^2 \int \frac{d |\bm{b}_T|}{(2\pi)} |\bm{b}_T|^3\, 
J_2(|\bm{b}_T|\,|\Phperp| ) \mN M_h z^2\ \tilde h_{1L}^{\perp a (1)}(x,z^2\bm{b}_T^2) \ \tilde H_1^{\perp a (1)}(z,\bm{b}_T^2)\ , 
\label{eq:c5}\displaybreak[0]\\
 F_{UT}^{\sin(3\phi_h-\phi_S)} &  = \xbj  \sum_a e_a^2 \int \frac{d |\bm{b}_T|}{(2\pi)} |\bm{b}_T|^4\, 
J_3(|\bm{b}_T|\,|\Phperp| )\frac{{\mN}^2 M_h z^3}{4} \tilde h_{1T}^{\perp a (2)}(x,z^2\bm{b}_T^2) \ \tilde H_1^{\perp a  (1)}(z,\bm{b}_T^2)\ .
\label{eq:Fstuctures}
\end{align}

\section{Cancellation of the soft factor in the Sivers asymmetry\label{appendix:derivation}}

Making use of the closure relation of the Bessel function
\begin{equation}
	\int_0^\infty d|\Phperp|\, |\Phperp|\, J_n(|\Phperp| \, |\bm{b}_{T}|)\, J_n(|\Phperp| \, \bpar) = \frac{1}{\bpar} \delta( |\bm{b}_T| - \bpar ) \, ,
\label{eq:closure}
\end{equation}
we  obtain for the expression in Eq.~\eqref{eq:ssa_sivers_denominator}
\bea
&& \int d |\Phperp|\, |\Phperp|\, d\phi_h\, d\phi_S\,  J_0( |\Phperp|\bpar  )\, \int \frac{d |\bt|}{(2\pi)} |\bt|  
J_{0} (|\bt| |\Phperp|)\, \FTStrufu_{UU,T} \nn
&&    = \xbj \sum_a e_a^2\ H_{UU,T}^{}(Q^2, \mu^2, \rho)\ \int d |\Phperp|\, |\Phperp|\, \int d\phi_h\, \int d\phi_S\, J_0(|\Phperp| \bpar) \nn
&& \hspace{.25cm}\times \int \frac{d |\bm{b}_T|}{(2\pi)}\, |\bm{b}_T|\   J_0(|\Phperp| \, |\bm{b}_{T}|)
\tilde f_1^{(0)a}(x,z^2 \bm{b}_T^2;\mu^2, \zeta,\rho)\ \tilde S^{(+)}(\bm{b}_T^2; \mu^2, \rho)\ \tilde D_1^{(0)a}(z,\bm{b}_T^2; \mu, \hat \zeta,\rho) \nn
&&  =2\pi \xbj\ 
\sum_a e_a^2\ H_{UU,T}^{}(Q^2, \mu^2, \rho) \  
 \tilde f_1^{(0)a}(x,z^2 \bpar^2;\mu^2, \zeta,\rho)\tilde S^{(+)}(\bpar^2; \mu^2, \rho) \tilde D_1^{(0)a}(z,\bpar^2; \mu, \hat \zeta,\rho)\nn
\label{eq:ssa_sivers_denominator_detail}
\eea
Next, we consider the following  expression in the numerator of the asymmetry, 
Eq.~\eqref{eq:ssa_sivers_numerator},
\bea
&&\int d |\Phperp| |\Phperp| \int d\phi_h \int d\phi_S
  \frac{2 J_1(|\Phperp| \bpar)}{\zh M \bpar}
\sin^2(\phi_h - \phi_S) 
\nonumber \\ && \qquad \qquad \qquad \qquad \times
\int \frac{d |\bm{b}_T|}{(2\pi)} |\bm{b}_T|^2 J_1(|\bm{b}_T|\,|\Phperp| ) 
 \FTStrufu_{UT,T}^{\sin(\phi_h-\phi_S)}\nn
&&=
\int d |\Phperp|\, |\Phperp|\, \int d\phi_h\, \int d\phi_S\, \frac{ 2 J_1(|\Phperp|\bpar)}{\zh M \bpar} \sin^2(\phi_h - \phi_S)
 \nonumber \\ && \qquad \times
 \xbj \sum_a e_a^2\ H_{UT,T}^{\sin(\phi_h-\phi_S)}(Q^2, \mu^2, \rho)\int \frac{d |\bm{b}_T|}{(2\pi)} |\bm{b}_T|^2\, J_1(|\bm{b}_T|\,|\Phperp| ) 
 \nonumber \\ && \qquad \times
 \mN z\tilde f_{1T}^{\perp (1)a}(x,z^2\bm{b}_T^2,\mu^2,\zeta,\rho)\ \tilde S^{(+)}(\bm{b}_T^2, \mu^2, \rho)\ \tilde D_1^{(0)a}(z,\bm{b}_T^2,\mu^2, \hat \zeta,\rho) \nn
 &&=
2\pi \xbj \sum_a e_a^2\ H_{UT,T}^{\sin(\phi_h-\phi_S)}(Q^2, \mu^2, \rho)  
 \tilde f_{1T}^{\perp (1)a}(x,z^2 \bpar^2,\mu^2, \zeta,\rho) 
 \nonumber \\ && \qquad \times
\tilde S^{(+)}(\bpar^2, \mu^2, \rho)  \tilde D_1^{(0)a}(z,\bpar^2,\mu^2, \hat \zeta/\zh,\rho) ,
\nn
\eea
where we have used the closure relation~Eq.(\ref{eq:closure}),  and 
\bea
\int_{0}^{2 \pi} \cos^2( m \, \phi_h + n \, \phi_S) d \phi_S = \int_{0}^{2 \pi} \sin^2( m \, \phi_h + n \, \phi_S) d \phi_S = \pi \; ,
\eea
for integer $n$ and $m$.  Thus, we obtain
\bea
&& A_{UT,T}^{\frac{2\, J_1(|\Phperp| \bpar)}{\zh M \bpar}
\sin(\phi_h - \phi_s)}(\bpar) 
 =  -2\frac{
\frac{\alpha^2}{y\slim Q^2}\, \frac{y^2}{(1-\varepsilon)}
\left( 1+\frac{\gamma^2}{2\xbj} \right)}
{\frac{\alpha^2}{y\slim Q^2}\, \frac{y^2}{(1-\varepsilon)}
\left( 1+\frac{\gamma^2}{2\xbj} \right)}
\nonumber \\
&& \hspace{0.25cm}\times\frac{\sum_a e_a^2H_{UT,T}^{\sin(\phi_h-\phi_S)}(Q^2, \mu^2, \rho) \tildeftperp(x,z^2 \bpar^2;\mu^2,\zeta,\rho)  \tilde S^{+}(\bpar^2,\mu^2, \rho) \tilde D_1^{(0)a}(z,\bpar^2; \mu^2, \hat \zeta,\rho)  
}{
\sum_a e_a^2 H_{UU,T}^{}(Q^2, \mu^2, \rho) \tilde f_{1}^{(0)a}(x,z^2 \bpar^2;\mu^2,\zeta,\rho) \tilde S^{+}(\bpar^2, \mu^2, \rho) \tilde D_{1}^{(0)a}(z,\bpar^2; \mu^2, \hat \zeta,\rho) 
} ,\nn
\label{eq:ssa_sivers_final}
\eea

\section{Bessel-weighted asymmetries \label{sec-bwa}}
Here we introduce the Bessel weights $w_n$ 
 \begin{equation}
	  w_n\equiv J_n(  |\Phperp| \bpar )\, n! \left(\frac{2}{\bpar}\right)^n
\; ,
\label{eq:weightfactorapp}
\end{equation}
and  summarize the Bessel-weighted asymmetries at leading twist:\\
\begin{align}
\hangalign \hspace{-3cm}  \underline{\rm Double\  Spin} & \nn
 A_{LL}^{{J}_0(|\Phperp| \bpar)}(\bpar) & =\  
2\frac{
\frac{\alpha^2}{y\slim Q^2}\, \frac{y^2}{(1-\varepsilon)}
\left( 1+\frac{\gamma^2}{2\xbj} \right)\sqrt{1-\varepsilon^2}}
{\frac{\alpha^2}{y\slim Q^2}\, \frac{y^2}{(1-\varepsilon)}
\left( 1+\frac{\gamma^2}{2\xbj} \right)}
\nonumber \\
&\hspace{-3cm} \times\frac{\sum_a e_a^2\,H_{LL}^{}(Q^2, \mu^2, \rho)\,  \tilde g_{1L}^{(0)a}(x,z^2 \bpar^2;\mu^2, \zeta,\rho)\,  \tilde D_1^{(0)a}(z,\bpar^2; \mu^2, \hat \zeta,\rho)  
}{
\sum_a e_a^2\, H_{UU,T}^{}(Q^2, \mu^2, \rho)\, \tilde f_{1}^{(0)a}(x,z^2 \bpar^2;\mu^2,\zeta,\rho)\,  \tilde D_{1}^{(0)a}(z,\bpar^2; \mu^2, \hat \zeta,\rho) 
}\;, 
\label{eq:ssa_dspin} \displaybreak[0] \\
\hangalign \hspace{-3cm}\underline{\rm Worm\ Gear} &  \nn
 A_{LT}^{\frac{2\, J_1(|\Phperp| \bpar)}{\zh M \bpar}\cos(\phi_h-\phi_S)}
(\bpar)  & =\  
2\frac{
\frac{\alpha^2}{y\slim Q^2}\, \frac{y^2}{(1-\varepsilon)}
\left( 1+\frac{\gamma^2}{2\xbj} \right)\sqrt{1-\varepsilon^2}}
{\frac{\alpha^2}{y\slim Q^2}\, \frac{y^2}{(1-\varepsilon)}
\left( 1+\frac{\gamma^2}{2\xbj} \right)}
\nonumber \\
& \hspace{-3cm}\times\frac{\sum_a e_a^2\,H_{LT}^{\cos(\phi_h-\phi_S)}(Q^2, \mu^2, \rho)\,  \tilde g_{1T}^{ (1) a}(x,z^2 \bpar^2;\mu^2, \zeta,\rho)\,  \tilde D_1^{(0)a}(z,\bpar^2; \mu^2, \hat \zeta,\rho)  
}{
\sum_a e_a^2\, H_{UU,T}^{}(Q^2, \mu^2, \rho)\, \tilde f_{1}^{(0)a}(x,z^2 \bpar^2;\mu^2,\zeta,\rho)\,  \tilde D_{1}^{(0)a}(z,\bpar^2; \mu^2, \hat \zeta,\rho) 
}\;,
\label{eq:ssa_worm} \displaybreak[0] \\
\hangalign \hspace{-3cm} \underline{\rm Collins} & \nn
A_{UT}^{\frac{2\, J_1(|\Phperp| \bpar)}{\zh M_h \bpar}\sin(\phi_h + \phi_s)}(\bpar) &  =\  
2\frac{
\frac{\alpha^2}{y\slim Q^2}\, \frac{y^2}{(1-\varepsilon)}
\left( 1+\frac{\gamma^2}{2\xbj} \right){\varepsilon}}
{\frac{\alpha^2}{y\slim Q^2}\,  \frac{y^2}{(1-\varepsilon)}
\left( 1+\frac{\gamma^2}{2\xbj} \right)}
\nonumber \\
& \hspace{-3cm}\times\frac{\sum_a e_a^2\,H_{UT}^{\sin(\phi_h+\phi_S)}(Q^2, \mu^2, \rho)\,  \tilde h_{1}^{(0)a}(x,z^2 \bpar^2;\mu^2, \zeta,\rho)\,  \tilde H_1^{\perp (1) a}(z,\bpar^2; \mu^2, \hat \zeta,\rho)  
}{
\sum_a e_a^2\, H_{UU,T}^{}(Q^2, \mu^2, \rho)\, \tilde f_{1}^{(0)a}(x,z^2 \bpar^2;\mu^2,\zeta,\rho)\,  \tilde D_{1}^{(0)a}(z,\bpar^2; \mu^2, \hat \zeta,\rho) 
}\;, \
\label{eq:ssa_collins} \displaybreak[0] \\
\hangalign \hspace{-3cm} \underline{\rm Boer-Mulders} & \nn
A_{UU}^{\frac{2\, J_2(|\Phperp| \bpar)}{\zh^2 M M_h\bpar^2}\cos(2\phi_h)}(\bpar)  &  =\  
2\frac{
\frac{\alpha^2}{y\slim Q^2}\, \frac{y^2}{(1-\varepsilon)}
\left( 1+\frac{\gamma^2}{2\xbj} \right){\varepsilon}}
{\frac{\alpha^2}{y\slim Q^2}\, \frac{y^2}{(1-\varepsilon)}
\left( 1+\frac{\gamma^2}{2\xbj} \right)}
\nonumber \\
& \hspace{-3cm} \times\frac{\sum_a e_a^2\,H_{UU}^{\cos(2\phi_h)}(Q^2, \mu^2, \rho)\,  \tilde h_{1}^{\perp (1) a}(x,z^2 \bpar^2;\mu^2, \zeta,\rho)\,  \tilde H_1^{\perp (1) a}(z,\bpar^2; \mu^2, \hat \zeta,\rho)  
}{
\sum_a e_a^2\, H_{UU,T}^{}(Q^2, \mu^2, \rho)\, \tilde f_{1}^{(0)a}(x,z^2 \bpar^2;\mu^2,\zeta,\rho)\,  \tilde D_{1}^{(0)a}(z,\bpar^2; \mu^2, \hat \zeta,\rho) 
}\;,
\label{eq:ssa_boermulders} \displaybreak[0] \\
\hangalign \hspace{-3cm} \underline{\rm Kotzinian-Mulders} & \nn
 A_{UL}^{\frac{2\, J_2(|\Phperp| \bpar)}{\zh^2 M M_h\bpar^2}\sin(2\phi_h)}(\bpar) &  =\  
2\frac{
\frac{\alpha^2}{y\slim Q^2}\, \frac{y^2}{(1-\varepsilon)}
\left( 1+\frac{\gamma^2}{2\xbj} \right){\varepsilon}}
{\frac{\alpha^2}{y\slim Q^2}\, \frac{y^2}{(1-\varepsilon)}
\left( 1+\frac{\gamma^2}{2\xbj} \right)}
\nonumber \\
& \hspace{-3cm}\times\frac{\sum_a e_a^2\,H_{UL}^{\sin(2\phi_h)}(Q^2, \mu^2, \rho)\,  \tilde h_{1L}^{\perp (1) a}(x,z^2 \bpar^2;\mu^2, \zeta,\rho)\,  \tilde H_1^{\perp (1) a}(z,\bpar^2; \mu^2, \hat \zeta,\rho)  
}{
\sum_a e_a^2\, H_{UU,T}^{}(Q^2, \mu^2, \rho)\, \tilde f_{1}^{(0)a}(x,z^2 \bpar^2;\mu^2,\zeta,\rho)\,  \tilde D_{1}^{(0)a}(z,\bpar^2; \mu^2, \hat \zeta,\rho) 
}\;,
\label{eq:ssa_kotzmulders} \displaybreak[0] \\
\hangalign \hspace{-3cm} \underline{\rm Pretzelosity} & \nn
 A_{UT}^{\frac{8\, J_3(|\Phperp| \bpar)}{\zh^3 M^2 M_h\bpar^3}\sin(3\phi_h - \phi_s)}(\bpar) &  =\  
2\frac{
\frac{\alpha^2}{y\slim Q^2}\, \frac{y^2}{(1-\varepsilon)}
\left( 1+\frac{\gamma^2}{2\xbj} \right){\varepsilon}}
{\frac{\alpha^2}{y\slim Q^2}\, \frac{y^2}{(1-\varepsilon)}
\left( 1+\frac{\gamma^2}{2\xbj} \right)}
\nonumber \\
& \hspace{-3cm}\times\frac{\sum_a e_a^2\,H_{UT}^{\sin(3\phi_h-\phi_S)}(Q^2, \mu^2, \rho)\,  \tilde h_{1T}^{\perp (2) a}(x,z^2 \bpar^2;\mu^2, \zeta,\rho)\,  \tilde H_1^{\perp (1) a}(z,\bpar^2; \mu^2, \hat \zeta,\rho)  
}{
\sum_a e_a^2\, H_{UU,T}^{}(Q^2, \mu^2, \rho)\, \tilde f_{1}^{(0)a}(x,z^2 \bpar^2;\mu^2,\zeta,\rho)\,  \tilde D_{1}^{(0)a}(z,\bpar^2; \mu^2, \hat \zeta,\rho) 
}\;.
\label{eq:ssa_pretz}
\end{align}

\section{Suppression of high transverse momentum}
\label{sec-highmomsuppr}

\subsection{Suppression of the tail in Bessel-weighted integrals}
\label{sec-tailintegrals}

In this section, we are interested in the contribution from the high-momentum region to Bessel-weighted integrals. 
We begin by deriving convergence criteria and upper bounds for an integral of the form 
\begin{equation}
	I_{n,A}(\xi,\Lambda_\omega) \equiv \int_{\Lambda_\omega}^\infty d\omega\,\omega\,J_n(\omega \xi)\,A(\omega)
	= \xi^{-2} \int_{\Lambda_\omega \xi}^\infty d\nu\,\nu\,J_n(\nu)\,A(\nu/\xi)\,.
	\label{eq-highmomtransf}
\end{equation}
In the subsections to follow, $\omega$ will assume the role of a momentum, $|\vprp{p}|$, $|\vprp{K}|$ or $|\Phperp|$, while $\xi$ will represent $|\vprp{\elll}|$ or $\bpar$. In the equation above, $A$ is a placeholder for a given function of $\omega$.
We restrict our discussion of the integral $I_{n,A}(\xi,\Lambda_\omega)$ to the region $\Lambda_\omega \xi \gg 1$, where the Bessel function in the integrand can be approximated by 
\begin{equation}
	J_n(\nu) \approx \sqrt{\frac{2}{\pi \nu}} \sin\left( \nu  + \frac{\pi}{4} - \frac{n\pi}{2} \right) \quad \text{for} \quad \nu \gg 1
	\label{eq-besselapprox}
\end{equation}
First, let us consider a function $A$ that fulfills the condition
\begin{equation}
	|A(\omega)| \leq c\, \omega^{-\alpha} \quad \text{for any}\quad  \omega \geq \Lambda_\omega
\end{equation}
where $c>0$ and $\alpha$ are two real valued constants.
Using the envelope of the Bessel function $|J_n(\nu)| \lesssim \sqrt{2/\pi \nu}$, we obtain
\bea
	\Big| I_{n,A}(\xi,\Lambda_\omega) \Big|
	&\leq & \xi^{-2} \int_{\Lambda_\omega \xi}^\infty d\nu\,\Big| \nu J_n(\nu) A(\nu/\xi) \Big| 
\nonumber \\ 
& \lesssim & \xi^{-2} \int_{\Lambda_\omega \xi}^\infty d\nu \sqrt{\frac{2\nu}{\pi}} c \frac{\xi^\alpha}{\nu^\alpha}  =  \frac{1}{\alpha-\frac{3}{2}} \sqrt{\frac{2 \Lambda_\omega^3}{\pi \xi}} c \Lambda_\omega^{-\alpha}  \sim  \xi^{-1/2},\nn
	\label{eq-tailAupperbound}
\eea
where the last equal sign only holds if the integral is convergent, i.e., if $\alpha > 3/2$.

Next, consider a function $B$ for which $\sqrt{\omega} B(\omega) $ is monotonously falling in the region $\omega \geq \Lambda_\omega$ and converging to zero for $\omega \rightarrow \infty$. 
In this case, we can make use of the oscillatory behavior of the Bessel function to show convergence of the integral $I_{n,B}(\xi,\Lambda_\omega)$. Let $N$ denote the smallest possible integer such that $\Lambda_\omega \xi + \pi/4 - n\pi/2 \leq N\pi$. We decompose the integration according to
\begin{align}
	I_{n,B}(\xi,\Lambda_\omega) & = 
	\underbrace{\xi^{-2} \int_{\Lambda_\omega \xi}^{N\pi -\pi/4 + n\pi/2} d\nu\ \nu\, J_n(\nu) B(\nu/\xi)}_{\displaystyle T_1}  
\nonumber \\ & +
	\underbrace{\xi^{-2} \sum_{j=0}^\infty \int_{(N+ j)\pi-\pi/4+n\pi/2}^{(N+j+1)\pi-\pi/4+n\pi/2} d\nu\ \nu\, J_n(\nu) B(\nu/\xi) }_{
	\displaystyle T_2}\ .
\end{align}
Applying the mean value theorem, we can find $\bar{\nu} \in [\Lambda_\omega \xi, N\pi - \pi/4 + 2\pi/2]$ such that
\begin{align}
	T_1& = \xi^{-2} \sqrt{\bar{\nu}} B(\bar{\nu}/\xi)  \int_{\Lambda_\omega \xi}^{N\pi -\pi/4 + n\pi/2} d\nu\ \sqrt{\nu}\, J_n(\nu)
\nonumber \\ &
\approx \xi^{-2} \sqrt{\bar{\nu}} B(\bar{\nu}/\xi) \sqrt{\frac{2}{\pi}}  \Big( \underbrace{\cos(\Lambda_\omega \xi + \pi/4 - n\pi/4)}_{\displaystyle \in [-1,1]} - \underbrace{\cos(N\pi)}_{\displaystyle (-1)^N} \Big) 
\end{align}
from which we derive bounds for the first term:
\begin{align}
	0 \ \leq\  (-1)^{N+1} T_1\  \leq\  2 \sqrt{\frac{2}{\pi}} \xi^{-2} \sqrt{\bar{\nu}} B(\bar{\nu}/\xi) 
	\ \leq\  2 \sqrt{\frac{2}{\pi}} \xi^{-2} \sqrt{\Lambda_\omega \xi + 2\pi} B(\Lambda_\omega) \ .
\end{align}
Using the mean value theorem again to determine the points $\bar{\nu}_j$, the second term becomes
\begin{align}
	T_2 & \approx \xi^{-2} \sum_{j=0}^\infty \sqrt{\bar{\nu}_j} B(\bar{\nu}_j/\xi) \int_{(N+ j)\pi-\pi/4+n\pi/2}^{(N+j+1)\pi-\pi/4+n\pi/2} d\nu\ \sqrt{\frac{2}{\pi}} \sin(\nu + \pi/4 -n\pi/2)
 \nonumber \\ &
=  2 \sqrt{\frac{2}{\pi}} \xi^{-2} \sum_{j=0}^\infty \sqrt{\bar{\nu}_j} B(\bar{\nu}_j/\xi) (-1)^{N + j}\, .
\end{align}
The expression on the right is an alternating series that fulfills the Leibnitz-test for convergence and is bounded by the size of its first term, 
\begin{align}
	0\ \leq (-1)^N T_2\ \leq\ 
	2 \sqrt{\frac{2}{\pi}} \xi^{-2} \sqrt{\bar{\nu}_0} B(\bar{\nu}_0/\xi) \leq\ 2 \sqrt{\frac{2}{\pi}} \xi^{-2} \sqrt{\Lambda_\omega \xi + 2\pi} B(\Lambda_\omega) . 
\end{align}
Since $T_1$ and $T_2$ have opposite signs, we arrive at a combined upper bound
\begin{align}
	\Big| I_{n,B}(\xi,\Lambda_\omega) \Big| \ \leq \ 2 \sqrt{\frac{2}{\pi}} \xi^{-2} \sqrt{\Lambda_\omega \xi + 2\pi}\, B(\Lambda_\omega)
	\ \approx \ 2 \sqrt{\frac{2 \Lambda_\omega}{\pi \xi^3}} \, B(\Lambda_\omega) \ \sim \ \xi^{-3/2}\, .
	\label{eq-tailBupperbound}
\end{align}

In summary, we find that the integral Eq.\ \eqref{eq-highmomtransf} converges for any function $A(\omega)$ that decays faster than $\omega^{-3/2}$ which in turn determines an upper bound of the integral of order $\xi^{-1/2}$,  Eq.\ \eqref{eq-tailAupperbound}.  The requirement for convergence can be relaxed to functions decaying faster than $\omega^{-1/2}$ if monotony of $\sqrt{\omega}$ times the function is ensured. In this case, Eq.\ \eqref{eq-tailBupperbound} gives an estimate of an upper bound that decays with $\xi^{-3/2}$.
We remind the reader that these bounds are only valid for $\xi \gg \Lambda_\omega^{-1}$, but for all $n$.

\subsection{Fourier-transformed \TMDPs, \TMDFs and their derivatives}

Using the mathematical results from the previous sub-section, 
we investigate which of the (derivatives of) Fourier-transformed \TMDPs $\tilde f^{(n)}(x,\vprp{\elll}^2)$ and \TMDFs $\tilde D^{(n)}(x,\vprp{\elll}^2)$ are well-defined by the right hand sides of Eq.\ \eqref{eq-btder}.
Their behavior  in the high transverse momentum region has been studied in detail in Ref. \cite{Bacchetta:2008xw}.
They find power-suppressed tails of the form 
\begin{equation}
	f(x,|\vprp{p}^2|) \sim \frac{1}{|\vprp{p}|^m} \times \text{``logarithmic modifications''}\, ,
\end{equation}
for integer powers $m$. Analogous expressions hold for the \TMDFs. 
Comparing the right hand side of Eq.\ \eqref{eq-btder} with the criterion for functions of type $B$, we find that convergence is maintained if $n < m - 1/2$. The logarithmic modifications do not play a significant role since logarithms grow more slowly than any polynomial.

The analysis of Ref.  \cite{Bacchetta:2008xw} reveals that (up to logarithmic modifications) $f_1,\allowbreak g_{1L},\allowbreak h_1,f^\perp,\allowbreak g_L^\perp,\allowbreak h_T,\allowbreak  h_T^\perp,\allowbreak f_T,\allowbreak g_T,\allowbreak h_L,\allowbreak h,\allowbreak e_L,\allowbreak e,\allowbreak f_L,\allowbreak g^\perp,\allowbreak e_T,\allowbreak e_T^\perp,\allowbreak D_1,\allowbreak D^\perp,\allowbreak G^\perp,\allowbreak H,\allowbreak E \sim 1/\vprp{p}^2$. For these functions, the corresponding zero-derivative and single-derivative Fourier-transforms $\tilde f^{(0)}(x,\vprp{\elll}^2)$, $\tilde f^{(1)}(x,\vprp{\elll}^2)$, $\tilde D^{(0)}(z,\vprp{\elll}^2)$ and $\tilde D^{(1)}(z,\vprp{\elll}^2)$ exist. A second group of distributions exhibits the high-momentum behavior $f_{1T}^\perp,\allowbreak  g_{1T},\allowbreak  h_{1L}^\perp,\allowbreak  h_1^\perp,\allowbreak  f_T^\perp,\allowbreak g_T^\perp,\allowbreak h_{1T}^\perp,\allowbreak H_1^\perp \sim 1/\vprp{p}^4$. For these latter functions, the existence of $n$-derivative Fourier-transforms $\tilde f^{(n)}(x,\vprp{\elll}^2)$ and $\tilde D^{(n)}(z,\vprp{\elll}^2)$ is ensured up to $n=3$. Again, we point out that these results are only valid for $|\vprp{\elll}|>0$, while the limiting case $|\vprp{\elll}| = 0$ leads to divergent integrals \cite{Bacchetta:2008xw}.

\subsection{Systematic errors from the region at large $\Phperp$}
\label{sec-Ytermsuppr}

TMD frameworks have been designed to give a good description of the cross section at low transverse momentum, i.e., for 
$|\Phperp|/z \ll Q$. However, in weighted asymmetries we integrate over the whole range of $|\Phperp|$.
The contributions from high $|\Phperp|$ thus lead to theoretical errors in the results if one does not have a description of the 
cross section that is valid there, even when one restricts to the region $z |\bm{b}_T| \gg 1/Q$. 
The $Y$ term can in principle be included to eliminate those errors, but its Fourier transform is expected to be power suppressed in the region $z |\bm{b}_T| \gg 1/Q$, 
because it was shown to be power suppressed at small $|\Phperp|$ \cite{Collins:1984kg,Collins:1981va}. Dropping the $Y$ term means that we approximate the full result by 
the large $|\Phperp|$-tail of the TMD expression. This in general may be a bad approximation, but the question is whether it will affect the result much for 
$z |\bm{b}_T| \gg 1/Q$. In addition, extending the integrals to arbitrarily large transverse momenta ignores the fact that the physical cross 
section should vanish above a certain maximum transverse momentum 
value $|\Phperp|_\text{max}$ (see also Refs. \cite{Ellis:1981sj,Collins:1981va}). In this appendix we are going to estimate the effect of these various simplifications. 

\begin{figure}[htb]
\begin{center}
	\includegraphics[width=0.6\textwidth]{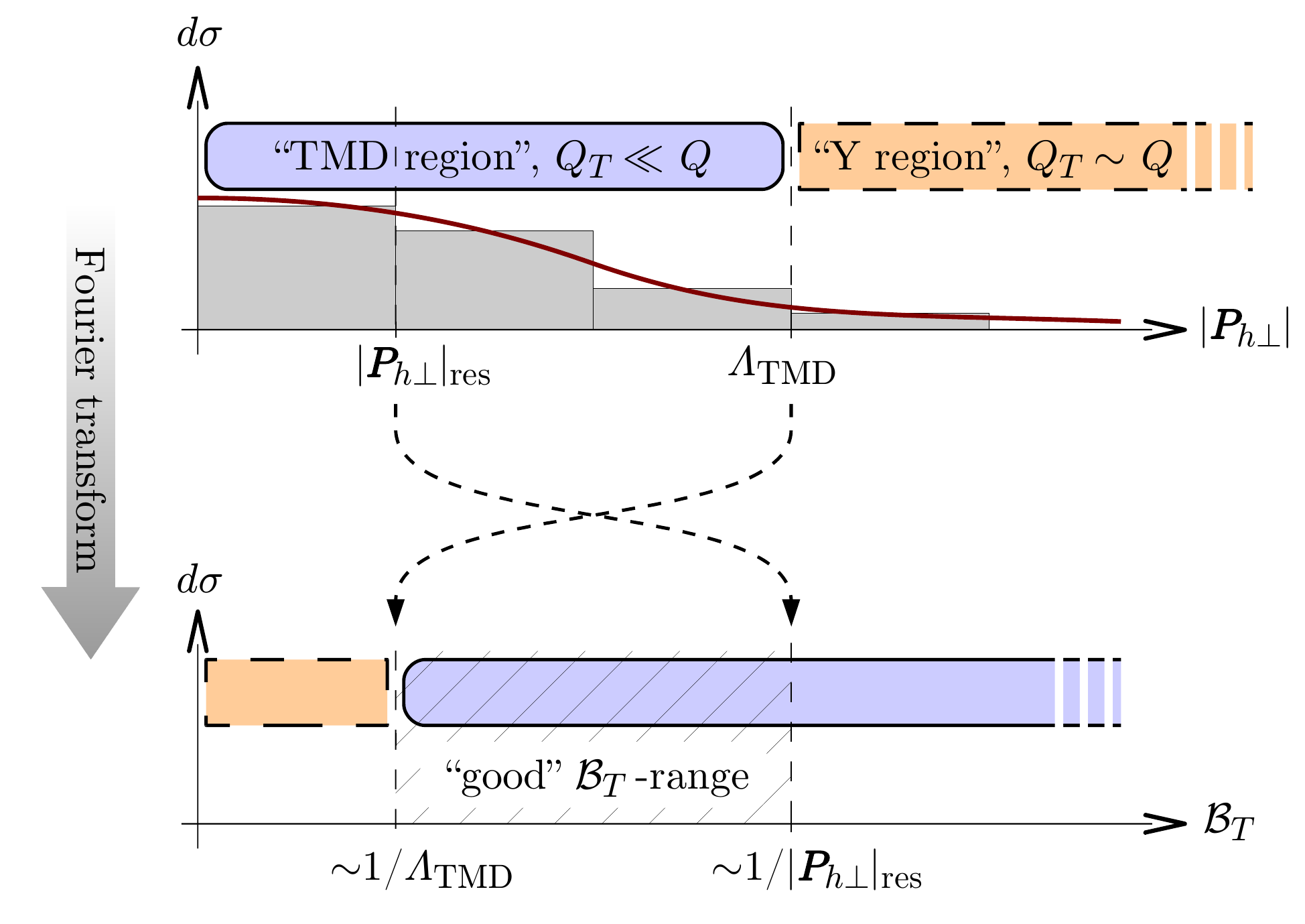}
	\caption{Schematic illustration of important scales for Bessel-weighted asymmetries before and after the Fourier-transform.\label{fig-TMDapprox}}
\end{center}
\end{figure}

The $Y$ term will be significant only in a finite region of $|\Phperp|$: between a scale $\Lambda_\text{TMD}$ and $|\Phperp|_\text{max}$. 
Note that both these scales will depend on $Q$. We can bound the error from neglecting the $Y$ term in terms of its maximal value. 
As long as $|\vprp{b}| \gg \Lambda_\text{TMD}^{-1} > |\Phperp|_\text{max}^{-1}$, we can approximate the Bessel-function 
as in Eq.\ \eqref{eq-besselapprox}  to obtain,
\begin{align}
\strudress{\tilde Y}(Q^2,\bm{b}_T^2) & \equiv \int  d|\Phperp|\ |\Phperp| \ 2\pi J_{N}( |\bt| |\Phperp|) \ \strudress{Y}(Q^2,\Phperp^2) \nonumber \\
& \hspace{-1cm}\approx \int_{\Lambda_\text{TMD}}^{|\Phperp|_\text{max}}  d|\Phperp|\ |\Phperp| \ 2\pi J_{N}( |\bt| |\Phperp|) \ \strudress{Y}(Q^2,\Phperp^2) \nonumber \\
	 & \hspace{-1cm}\lesssim\ \left(  |\Phperp|_\text{max} - \Lambda_\text{TMD} \right)\ 2\sqrt{\frac{2\pi}{|\vprp{b}| \Lambda_\text{TMD} }}\ \Big| \strudress{Y} \Big|_\text{max}\ .
	 \label{eq-Ybound}
\end{align}
Here $| \strudress{Y} |_\text{max}$ is the maximum absolute value of $Y$ in the range between $\Lambda_\text{TMD}$ and $|\Phperp|_\text{max}$. It can be estimated from the (perturbatively calculable) $Y$-term. Thus, Eq.~\eqref{eq-Ybound} shows that the theoretical error from neglecting the $Y$ term is (at least) suppressed as $|\vprp{b}|^{-1/2}$. An explicit treatment of the $Y$-term in Eq.\ \eqref{eq:strucNLO} could eliminate this theoretical error to a given order in $\alpha_s$ in the Fourier transformed \TMDPs and \TMDFs extracted using Bessel weighting. We will not do this here. 

The second error coming from extending the TMD expression beyond $|\Phperp|_\text{max}$ is more suppressed and therefore
less of a concern. Following a similar procedure as before we can estimate it to be suppressed as $|\vprp{b}|^{-3/2}$. Let $[\strudress{F}]_\text{TMD}$ denote the structure functions as determined purely within the TMD framework, i.e., from convolutions of \TMDPs, \TMDFs and a potential soft factor. The contribution to its Fourier transform coming from the large $|\Phperp|$ region can be bounded using that the TMD expression (times $|\Phperp|^{1/2}$) is a monotonically decreasing function of $|\Phperp|$. Thus,  applying  Eq.\ \eqref{eq-tailBupperbound},
\begin{align}
 \int_{|\Phperp|_\text{max}}^{\infty}  d|\Phperp|\ |\Phperp|\ 2\pi J_{N}( |\bt| |\Phperp|) \ [\strudress{F}]_\text{TMD}(Q^2,\Phperp^2) 
\nonumber \\
	\lesssim\ 4 \sqrt{ \frac{2\pi|\Phperp|_\text{max}}{|\vprp{b}|^3}}\ \Big| [\strudress{F}]_\text{TMD}(Q^2,|\Phperp|_\text{max}^2) \Big|\ ,
	 \label{eq-TMDbound}	
\end{align} 
where the upper bound applies as long as $|\vprp{b}| \gg |\Phperp|_\text{max}^{-1}$. This second error is therefore far less important 
than neglecting the $Y$ term. The reason this same behavior could not be obtained for the $Y$ term is that it is not expected to
be a monotonically falling function of $|\Phperp|$. 

Finally, let us consider what error would be introduced if all $|\Phperp|$ integrations of the experimental data were to be cut off at $\Lambda_\text{TMD}$. In this case, we would be able to use Eq.\ \eqref{eq-TMDbound} as an error estimate, except that $ |\Phperp|_\text{max}^{-1} $ would need to be replaced by $\Lambda_\text{TMD}$. Again the error estimate would be valid provided $|\vprp{b}| \gg \Lambda_\text{TMD}^{-1}$ and provided the structure function times $|\Phperp|^{1/2}$ is monotonically falling, i.e., in its tail region, beyond $\Lambda_\text{TMD}$. This simple cutoff method is expected to be useful when $Q^2$ is very large, such that $\Lambda_\text{TMD}$ can be chosen large with confidence.

In Fig.\ \ref{fig-TMDapprox} we illustrate how the contributions from the TMD and the $Y$ dominated regions contribute to 
the Fourier transform. The contributions from the region $|\Phperp|>\Lambda_\text{TMD}$ are only suppressed in the region of large ${\cal B}_T > 1/\Lambda_\text{TMD}$. Therefore, an analysis without $Y$ term at too low values of ${\cal B}_T$ has to be considered with caution. However, 
also the region of large ${\cal B}_T$ has to be treated with care in case of Bessel weighting, as one starts to probe the oscillations of the Bessel function. This is relevant whenever $1/\bpar$ becomes smaller than the experimental resolution in transverse momentum. 
A finite transverse momentum resolution $|\Phperp|_\text{res}$ can, for example, be a result of binning of the experimental data, as indicated in the figure.

\bibliography{biblio-2}

\end{document}